\documentclass[10pt,prx,twocolumn,amsmath,amssymb,floatfix,notitlepage]{revtex4-2}
\usepackage[T1]{fontenc}
\usepackage[utf8]{inputenc}
\usepackage{lmodern}
\usepackage{microtype,bm,bbm}
\usepackage{graphicx,booktabs}
\usepackage{times}
\usepackage[usenames,dvipsnames]{xcolor}
\usepackage[english]{babel}
\usepackage{upgreek}
\usepackage{braket}
\usepackage{subfigure}
\usepackage{dsfont}
\usepackage{transparent}

\usepackage{hyperref}
\hypersetup{
  colorlinks,
  allcolors=NavyBlue,
  linktoc=all
}

\usepackage[capitalize]{cleveref}
\crefformat{section}{Sec.~#2#1#3}

\newcommand{\id}{\mathbbm{1}}

\renewcommand*\vec[1]{\mathbf{#1}}
\renewcommand{\hbar}{}

\begin{document}

\title{Quantum simulation with fully coherent dipole--dipole-interactions\\ mediated by three-dimensional subwavelength atomic arrays}
\author{Katharina Brechtelsbauer$^{1,2}$}
\author{Daniel Malz$^{1,3}$}
\affiliation{$^1$ Max Planck Institute of Quantum Optics, Hans-Kopfermann-Stra{\ss}e 1, D-85748 Garching, Germany}
\affiliation{$^2$ Department of Physics, Technical University of Munich, James-Franck-Straße 1, D-85748  Garching,  Germany}
\affiliation{$^3$ Munich Center for Quantum Science and Technology, Schellingstra{\ss}e 4, D-80799 München, Germany}
\date{\today}
\pacs{}

\begin{abstract}
Quantum simulators employing cold atoms are among the most promising approaches to tackle quantum many-body problems. Nanophotonic structures are widely employed to engineer the bandstructure of light and are thus investigated as a means to tune the interactions between atoms placed in their vicinity.
A key shortcoming of this approach is that excitations can decay into free photons, limiting the coherence of such quantum simulators.
Here, we overcome this challenge by proposing to use a simple cubic three-dimensional array of atoms to produce an omnidirectional bandgap for light and show that it enables coherent,  dissipation-free interactions between embedded impurities.
We show explicitly that the band gaps persist for moderate lattice sizes and finite filling fraction, which makes this effect readily observable in experiment.
Our work paves the way toward analogue spin quantum simulators with long-range interactions using ultracold atomic lattices, 
and is an instance of the emerging field of atomic quantum metamaterials.
\end{abstract}
\maketitle

\section{Introduction}
The possibility of engineering and manipulating interactions between atoms is an essential requirement for realizing analogue quantum simulators~\cite{Noh2017}.
Recent theoretical and experimental approaches use photonic crystal waveguides to manipulate the electromagnetic environment of individual atoms~\cite{Gonzalez-Tudela2015,Douglas_2015,Hood10507,liu2017,Chang2018,Yu2019}.
In particular, photonic crystal waveguides can host bandgaps, such that quantum emitters with transition frequencies in the bandgap cannot decay into the waveguide and instead form exponentially localized atom--photon-bound states~\cite{PhysRevX.6.021027,PhysRevA.93.033833}.
This mechanism can be used to mediate interactions of tuneable range~\cite{de_Vega_2008}.
With control over the emitter spacing and the nature of their coupling to electromagnetic modes, a wide class of quantum spin models can be engineered~\cite{Gonzalez-Tudela2015,Douglas_2015}, which constitutes a highly promising avenue for cold-atom simulators.

To achieve high coupling strengths, atoms have to be trapped at subwavelength distances from the nanophotonic structures, which has proven very challenging. This has motivated a number of proposals and experimental advances~\cite{Gonzalez-Tudela2015,Yu2019,Beguin2020}.
Very recently, it has been shown that one- and two-dimensional atomic arrays can emulate nanophotonic structures and can be employed to control linewidth and dipole--dipole-interactions of additional impurity atoms
\cite{masson2019atomicwaveguide,patti2020controlling}.
Without fabrication disorder and surface Casimir forces, atomic arrays promise to simplify trapping of impurities close by and may yield more homogeneous systems.
Indeed, it is known that dense, ordered arrays may have rich bandstructures~\cite{Bettles2017,Perczel2017,Perczel2017a} and optical properties~\cite{Jenkins2013,Bettles2016,Shahmoon2017,Asenjo-Garcia2017,Shahmoon2020,Rui2020}.
Yet they come with the disadvantage that in optical dipole traps it is challenging to achieve highly subwavelength trapping, which is required since otherwise the typcial atom--atom interaction strength is comparable to their free-space decay rate~\cite{masson2019atomicwaveguide,patti2020controlling}.
 
The competition between unitary evolution and dissipation arises as all the approaches above feature one- and two-dimensional photonic nanostructures, which leave a third dimension into which photons can decay.
This casts serious doubt on the prospect of high-fidelity quantum simulation with nanophotonic structures.
Restricting solid-state nanophotonic structures to two dimensions is natural due to fabrication constraints (although note Ref.~\cite{3Dnanophotonic}) and because implanting quantum emitters comes with other challenges such as non-radiative decays, inhomogeneous broadening, and disorder in their positions~\cite{Bradac2019}.

In this paper, we thus propose to use three-dimensional atomic arrays to engineer omnidirectional bandgaps and furthermore to mediate interactions between impurity atoms.  
In the past, the occurence of bandgaps in atomic arrays has been discussed in several works~\cite{fluctuations,Klugkist2006,PhysRevA.52.1394,PhysRevLett.77.2412,PhysRevA.84.043833}.
However, the diamond lattice so far is the only known atomic array that can host an omnidirectional band gap~\citep{antezza}. Here, we show that bandgaps for one and both polarizations of light can be opened in simple cubic lattices by applying a suitable magnetic field and AC Stark shifts, compatible with current state-of-the-art experiments. 
We provide analytical insight in the nature and size of the band gap, which we verify with numerical simulation for both finite and infinite lattices. 

\begin{figure}[tb]
\includegraphics[width=1.04 \linewidth]{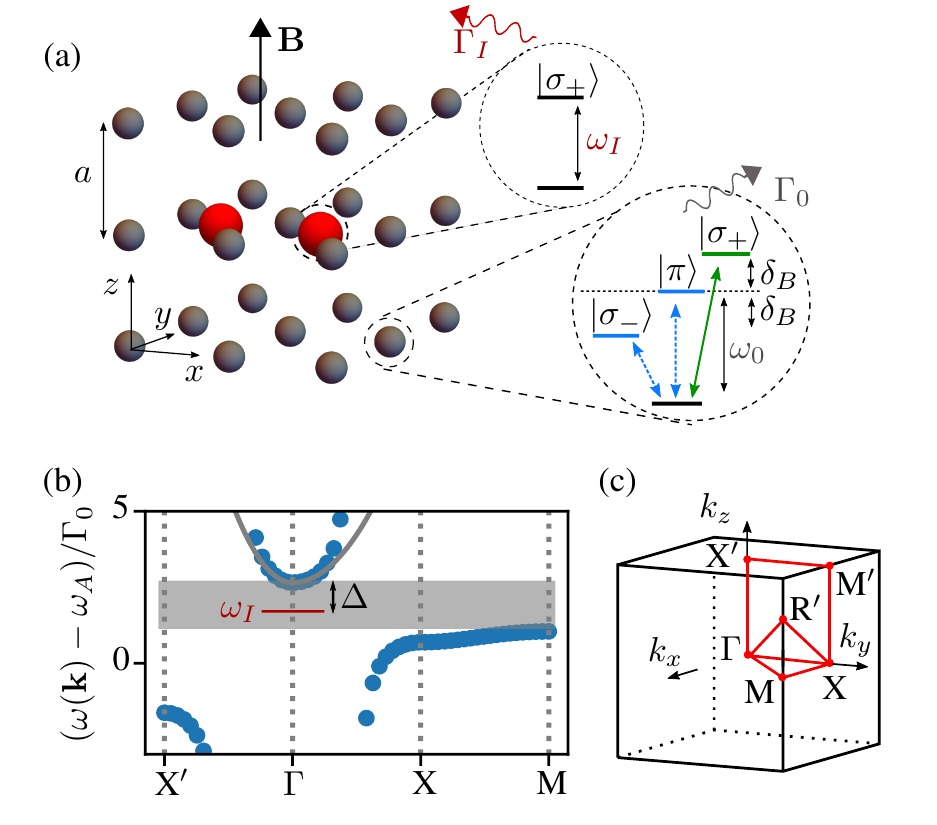}
\caption{\textbf{(a) Sketch of the proposed setup.} A three-dimensional array of atoms plays the role of a nanophotonic structure that modifies the band structure of light. Impurity atoms placed within the array interact via tunable array modes. If the array has an omnidirectional bandgap, decay of the impurities is suppressed and they undergo purely Hermitian dynamics. For the array we consider two-level atoms (green (solid) transition) as well as four-level atoms (green (solid) + blue (dashed) transitions). (b) The resulting bandstructure for $\sigma_+$-polarized two-level array atoms, exhibiting a band gap. The frequency of the impurity atoms $\omega_I$ can be tuned into the bandgap using Raman transitions.  (c) Path in the Brillouin zone corresponding to the plot in (b).}
\label{setup}
\end{figure}

We argue analytically and demonstrate numerically that our setups indeed can be used to mediate tunable long-range interactions between embedded impurity atoms. In the limit of infinite system size we prove that impurity atoms have infinite lifetime if their transition frequency lies in the band gap.
Analyzing finite-size effects, we show that the impurity decay rate decreases exponentially with system size and conclude that our proposal still works for lattice sizes of $20\times20\times20$-atoms, which is readily realized in experiment~\cite{Greiner2002,PhysRevLett.94.080403}.
By exponentially enhancing the coherence of light-based quantum simulators, our work removes an otherwise fundamental limitation for this class of quantum simulators.

The rest of this article is organized as follows. In Sec.~\ref{model} we introduce the theoretical framework for atom-light interactions. Within this model we investigate how one or both polarizations of light can be gapped in Sec.~\ref{photonicbandgaps}. In Sec.~\ref{mediatingimpurityinteractions} we show that these setups can be used to modify dipole--dipole interactions and the linewidth of impurity atoms. We discuss potential implementations in Sec.~\ref{implementation} and the application of our scheme to quantum simulation in Sec.~\ref{quantumsimulation}. We conclude in Sec.~\ref{conclusion}.

\section{Model}
\label{model}

We consider three-dimenional simple cubic arrays of two-level atoms as well as arrays of four-level atoms, which have one ground state and three excited states with excitation energies $\omega_0$.   
In the dipole approximation the light-matter Hamiltonian describing the interactions of the atoms with the radiation field is given by~\cite{Lehmberg1970,Lehmberg1970a}
\begin{align}
H_\text{lm}&=\sum_{\mathbf{R}_i}\sum_{\alpha}  \omega_0  \hat{b}^\dagger_{i,\alpha} \hat{b}_{i,\alpha}+\sum_{ \mathbf{k}}\sum_{\boldsymbol{\epsilon} \perp \mathbf{k}} c k \hat{a}^\dagger_{\mathbf{k},\boldsymbol{\epsilon}} \hat{a}_{\mathbf{k},\boldsymbol{\epsilon}}\notag \\&-\sum_{\mathbf{R}_i}\sum_{\alpha}  \hat{\mathbf{D}}_{i\alpha} \cdot \hat{\mathbf{E}}(\mathbf{R}_i).
\label{hamiltonian}
\end{align}
Here the sum over $\alpha$ runs over all excited states (\emph{cf.} Fig.~\ref{setup}) and $\vec R_i$ are the sites of the three-dimensional lattice. We represent the array atoms by bosonic annihilation operators $\hat{b}_{i,\alpha}$, which is valid if the density of excitations is low. Assuming that the strength of the dipoles is the same for all excited states, the dipole operator of the atom at position $\mathbf{R}_i$ is given by $\hat{\mathbf{D}}_{i\alpha}=d_0(\mathbf{d}_\alpha\hat{b}^\dagger_{i,\alpha}+\mathbf{d}_\alpha^*\hat{b}_{i,\alpha})$, where $\mathbf{d}_{\alpha}$ and $d_0$ are the direction and the strength of the dipole moment associated with the transition from the ground state to the excited state $\alpha$. The photons are described by  the photon annihilation operators $\hat{a}_{\mathbf{k},\boldsymbol{\epsilon}}$ and the electric field operator is
\begin{align}
\hat{\mathbf{E}}(\mathbf{r})=\sum_{\mathbf{k}}\sum_{\boldsymbol{\epsilon} \perp \mathbf{k}}\left(\sqrt{\frac{\hbar c k}{2\epsilon_0 V}} \boldsymbol{\epsilon} \hat{a}_{\mathbf{k},\boldsymbol{\epsilon}} e^{i\mathbf{k}\mathbf{r}} + \text{h.c.}\right).
\end{align} 
Eliminating the photons adiabatically via the Born-Markov approximation, the atom dynamics can be described by a non-Hermitian effective Hamiltonian of the form~\cite{Lehmberg1970,Lehmberg1970a} 
\begin{align}
H_\text{array}&= \sum_{\mathbf{R}_i}\sum_{\alpha} \hbar \left(\omega_\text{A}-i\frac{\Gamma_0}{2}\right)  \hat{b}^\dagger_{i,\alpha} \hat{b}_{i,\alpha} \notag\\&+\frac{3 \pi c \Gamma_0 }{\omega_0} \sum _{\mathbf{R}_i\neq \mathbf{R}_j}  \sum_{\alpha,\beta} \mathbf{d}_\alpha^* \cdot \boldsymbol{\mathcal{G}}(\mathbf{R}_i-\mathbf{R}_j)\cdot \mathbf{d}_\beta \  \hat{b}_{i,\alpha}^\dagger \hat{b}_{j,\beta},
\label{fullheff}
\end{align} 
where $\Gamma_0=(d_0^2\omega_0^3)/(3\pi \epsilon_0 c^3)$ is the  emission rate and $\omega_A$ deviates from $\omega_0$ by Lamb shift-type terms. In free space the Green's tensor $\boldsymbol{\mathcal{G}}(\mathbf{r})$ is given by the dyadic Green's function~\cite{PhysRevA.57.3931}. 
Note that the effective Hamiltonian in \cref{fullheff} describes the dynamics of the single-excitation sector completely if there is no driving field, such that quantum jumps can be neglected~\cite{PhysRevA.96.063801}.
For an infinite periodic lattice it is convenient to use Bloch's theorem to simplify the Hamiltonian. For two-level atoms with dipole moment $\mathbf{d}$ one directly obtains the dispersion relation
\begin{align}
\omega(\mathbf{k})-i\gamma(\mathbf{k})/2
&=\omega_\text{A}-i\Gamma_0/2+\frac{3\pi c\Gamma_0}{\omega_0} \mathbf{d}^* \cdot \tilde{\boldsymbol{\mathcal{G}}}(\mathbf{k}) \cdot \mathbf{d},
\label{onepoldispersion}
\end{align}
while for four-level atoms the problem of finding the eigenvalues of $H_\text{array}$ reduces to diagonalizing a $3\times 3$-matrix of the form
\begin{align}
\mathbf{M}=(\omega_A-i\Gamma_0/2)\mathds{1}+\frac{3 \pi c \Gamma_0}{\omega_0}\tilde{\boldsymbol{\mathcal{G}}}(\mathbf{k}),
\label{main}
\end{align}
where the atom--atom interactions are given by the discrete Fourier transform of the Green's tensor $\tilde{\boldsymbol{\mathcal{G}}}(\mathbf{k})=\sum_{\mathbf{R} \neq 0} \exp({-i\mathbf{k}\mathbf{R}})\boldsymbol{\mathcal{G}}(\mathbf{R}) $. 

Note that despite appearances, $\mathbf{M}$ is Hermitian and therefore only has real eigenvalues (see Eq.~\ref{calcdftofg}), such that all eigenstates of an infinite three-dimensional atomic array have infinite lifetime~\cite{fluctuations}.
In the above expression, $i\Gamma_0/2$ is cancelled by the non-Hermitian part in $\tilde{\boldsymbol{\mathcal{G}}}(\mathbf{k})$.

\section{Photonic band gaps}
\label{photonicbandgaps}
In this section we determine under which circumstances the model outlined above predicts photonic band gaps. We first show that a simple cubic lattice of two-level atoms opens a gap for light whose polarization coincides with the polarization of the atomic transition.
We then show that omnidirectional band gaps for both polarizations of light can be opened with four-level atoms and a suitable combination of magnetic fields and AC Stark shifts.

\subsection{Band gap for circularly polarized light}
\label{gapone}
We consider a simple cubic array of two-level atoms with circular polarization, which gaps out light of the same polarization, provided the array spacing $a$ fulfils the subwavelength condition $a<\lambda/2$, where $\lambda$ is the wavelength of light.

\begin{figure}[t]
\includegraphics[width=1.05 \linewidth]{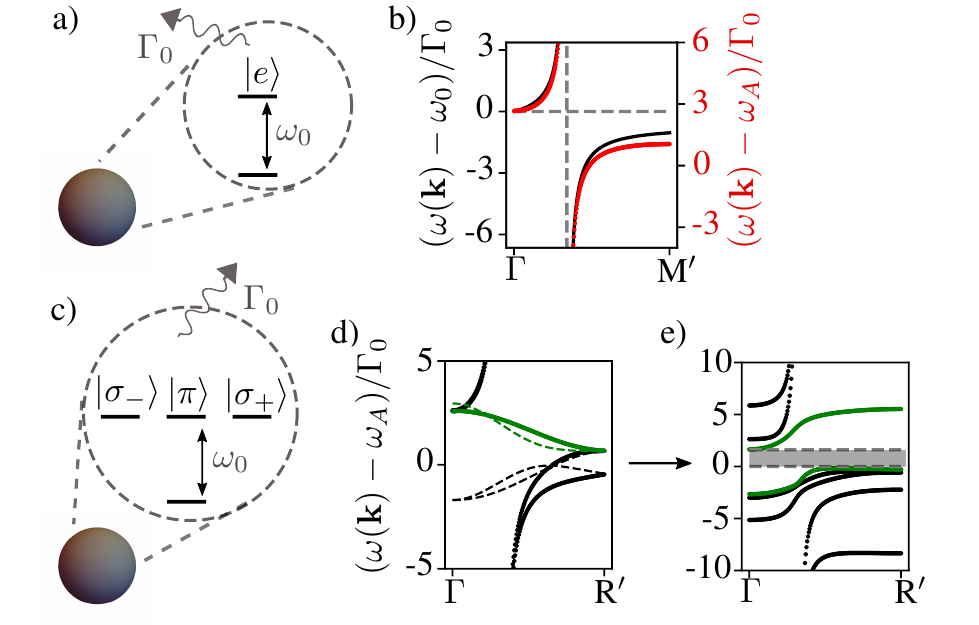}
\caption{\textbf{Bandstructure of light} interacting with two- (a-b) and four-level atoms (c-e). $\Gamma$, M$^\prime$ and R$^\prime$ are defined in \cref{setup}. (b) The black curves show the dispersion when the $\mathbf{G} \neq 0$-bands are neglected and the rotating-wave approximation is applied. The red (gray) points show the photonic dispersion given by Eq.~(\ref{onepoldispersion}). The dashed lines are the bands in the absence of coupling. One effect of the coupling to higher photon bands is that the atomic frequency $\omega_0$ is shifted to $\omega_A$, which is why the scale for red points is shifted with respect to the black. In the case of two-level atoms, the interactions with higher photon bands only enter as small perturbations. (d) In the presence of three dipoles the bandstructure is not gapped due to symmetry. The two bands coloured in green (gray) touch at R' and can be gapped by introducing a suitable perturbation as discussed in the main text (e). Note that the two highest black bands in (d) are degenerate.}
\label{gappingmechanism}
\end{figure}

Due to the periodicity of the array, the atoms couple to infinitely many photon bands $\omega_\mathbf{G}(\mathbf{k})=c|\mathbf{k}-\mathbf{G}|$, where $\mathbf{G}$ are reciprocal lattice vectors.
In this simple case, we can restrict our attention to the lowest photon bands ($\mathbf{G}=0$) as the gap opens generically due to hybridization of the atom band with those modes. The coupling to higher bands only yields a small shift, which is illustrated in \cref{gappingmechanism}(b).
Thus, in the rotating-wave approximation and neglecting higher bands, the Hamiltonian in \cref{hamiltonian} can be approximated by
\begin{align}
H_\text{lm}\approx\sum_{\mathbf{k} \in 1.\text{Bz.}} \begin{pmatrix} b_\mathbf{k}^\dagger \\ a_{\mathbf{k}\boldsymbol{\epsilon}_1}^\dagger\\ a_{\mathbf{k}\boldsymbol{\epsilon}_2}^\dagger\end{pmatrix}^\top
\begin{pmatrix} \omega_0 & g_1 \sqrt{k} & g_2  \sqrt{k}\\ g_1  \sqrt{k} & ck & 0\\ g_2  \sqrt{k} &0&ck \end{pmatrix}
\begin{pmatrix} b_\mathbf{k} \\ a_{\mathbf{k}\boldsymbol{\epsilon}_1}\\ a_{\mathbf{k}\boldsymbol{\epsilon}_2}\end{pmatrix},
\label{eqapproxlm}
\end{align}
where the coupling is defined as $g_i=d_0\sqrt{c/(2\epsilon_0 V_L)} \mathbf{d} \cdot \boldsymbol{\epsilon}_i,$ for $i \in \{1,2\}$, where $V_L=a^3$ is the volume of the unit cell and $b_\mathbf{k}=1/\sqrt{N^3} \sum_j \exp{(-i\mathbf{k}\mathbf{R}_j)}b_j$.
We can always choose a basis such that at least one polarization of light is orthogonal to the atomic polarization and thus decoupled.
Without loss of generality we thus choose $g_2=0$, such that finding the eigenvalues of \cref{eqapproxlm} reduces to diagonalizing a $2\times2$-matrix.
For $g_1=0$ the eigenvalues of this matrix cross at $k_0=\omega_0/c$. By coupling the levels $g_1 \neq 0$ the  crossing is avoided, such that a gap of width $\propto g_1^2$ opens up near $\omega_0$. The gap closes  for $\mathbf{k} \parallel \mathbf{d}$. This can be prevented if two components of $\mathbf{d}$ differ by a complex phase, as in circular polarization. The nature of this omnidirectional bandgap is subtle since the polarization of the gapped mode depends on the wavevector $\mathbf{k}$. However, a $\sigma_+$-polarized impurity atom placed in an array of $\sigma_+$-polarized atoms ``experiences'' an omnidirectional bandgap.

Note that there is a subtlety hidden in the arguments above.
The problem is that we have neglected standing waves of the electromagnetic field with nodes at the lattice sites, commonly referred to as the free-photon modes~\cite{Klugkist2006}.
Such modes would couple to impurities hosted within the lattice and thus should be avoided.
As they need to have a node at each lattice site, these modes have a minimum wavevector that depends on lattice geometry. 
In the simple cubic lattice the cutoff is $k_c=\pi/a$, which means that the lowest energy at which these modes appear is $\omega_c=c\pi/a$.
To shift these modes away from resonance, we require the subwavelength condition $a<\lambda_0/2$.

\subsection{Omnidirectional bandgap for both polarizations}

To open an omnidirectional band gap for both polarizations of light, we need to move to four-level atoms. 
While this alone is not sufficient to produce a bandgap,
we find that through judicious choice of magnetic field, as well as AC and DC Stark shifts, an omnidirection gap for both polarizations can be produced.
In particular, we consider a situation sketched in \cref{setup} (without impurity atoms), with the following ingredients:
\begin{itemize}
    \item[(i)] four-level atoms corresponding to a $J=0$ to $J=1$ transition, \emph{e.g.}, in $^{84}$Sr,
    \item[(ii)] a homogeneous magnetic field applied in the $z$-direction,
    \item[(iii)] two off-resonant lasers with wave vectors $\mathbf{k}=(k_x,k_y,\pm\pi/(2a))$ and polarizations $\boldsymbol{\epsilon}\propto(e^{i\phi}/\sqrt{2},i e^{i\phi}/\sqrt{2},\mp1)$ to produce Stark shifts of the $\sigma_+$- and $\pi$-transition of every second layer of atoms \footnote{$\phi,k_x$ and $k_y$ are determined by the condition $\boldsymbol{\epsilon} \perp \mathbf{k}$ and the detuning of the laser from $\omega_0$.}, and
    \item[(iv)] two global laser fields (of different strength) to produce Stark shifts of the $\pi$- and $\sigma_-$ transition.
\end{itemize}
In the rest of this section, we detail why these requirements arise and how they contribute to the band gap.

We note that we need to consider three dipole transitions, because for just two transitions
one can always find a $\mathbf{k}$ such that
$\mathbf{d}_2^* \cdot \boldsymbol{\epsilon}_2\propto\mathbf{d}_2^* \cdot(\mathbf{d}^* \times \mathbf{k})=  \mathbf{k} \cdot (\mathbf{d}_2^* \times \mathbf{d}^*)=0,$
independent of the dipole moment $\mathbf{d}_2$.
Hence we need all three orthogonal polarizations.
Surprisingly, even with all three polarizations, no gap opens, as shown in \cref{gappingmechanism}. Indeed, the two lowest photon modes ($\mathbf{G}=0$) both couple to an atomic polarization and are thus gapped. However there is an additional band (green band in \cref{gappingmechanism}), which closes the gap, since at some points at the edge of the Brillouin zone, this band and at least one of the other bands have to be degenerate due to symmetry~\cite{antezza,fluctuations,Klugkist2006,PhysRevA.84.043833}. In the simple cubic lattice these degeneracy points are the edges of the cube describing the Brillouin zone, which are defined by
$|k_x|=|k_y|=\pi/a $, $|k_x|=|k_z|=\pi/a$ and $|k_y|=|k_z|=\pi/a$.

\begin{figure}[tb]
\includegraphics[width= \linewidth]{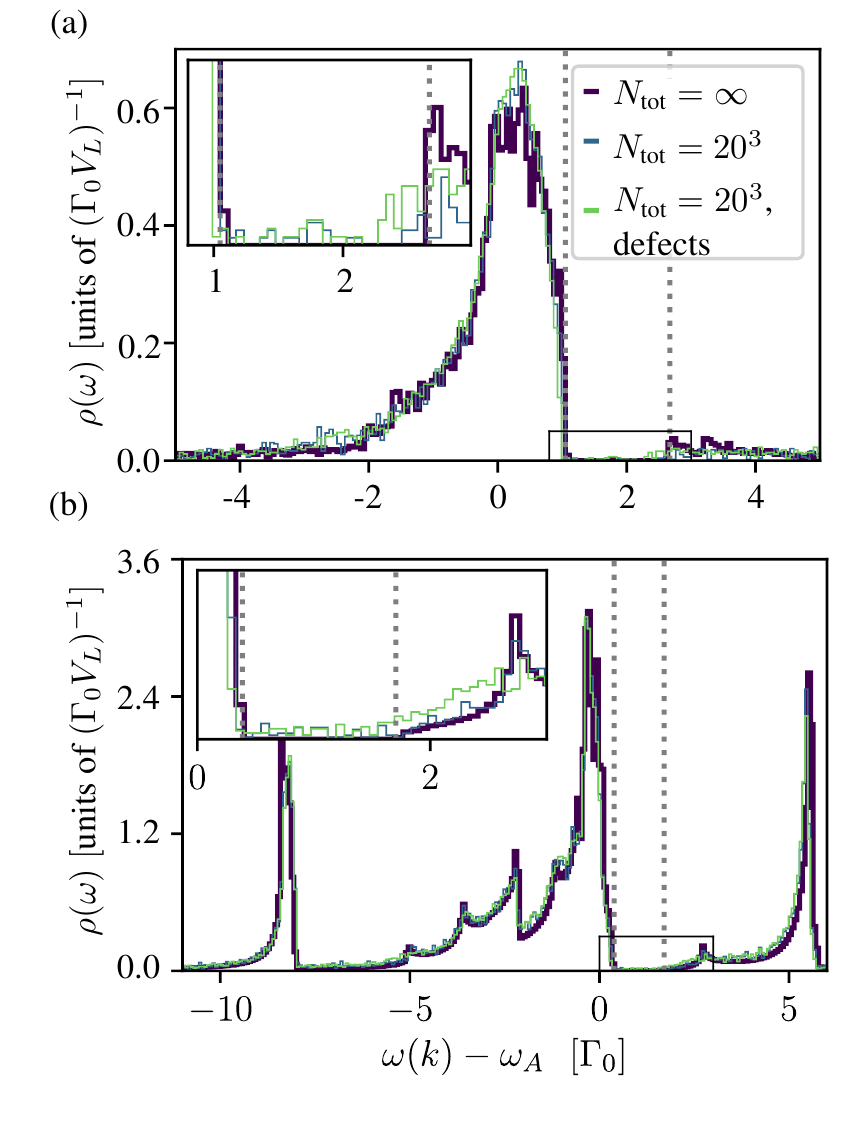}
\caption{\textbf{Density of states} with an omnidirectional band gap for one (a) and both (b) polarizations for $a=0.24 \lambda_0$. In panel (a) we consider an array of $\sigma_+$-polarized atoms. In the infinite case (blue) a gap arises (dashed lines). The inset is a magnification of the gapped region. In the finite case (orange) some states in the gap appear. If furthermore some defects are taken into account (green), additional states appear in the gap. We assume here a defect density $N_\mathrm{def}/N_\mathrm{tot}=0.1$~\cite{Rui2020}, where $N_\mathrm{tot}=N\times N\times N=20^3$ is the number of lattice sides. The energies of the infinite system were calculated with a mesh of $10^6$ points  for $k_x,k_y,k_z\geq 0$. In panel (a) the histograms contain 164 bins and for panel (b) the number of bins is 237. Other relevant parameters for (b) are $a_\text{ho}=0.09a$, $\delta_B=\mu_B B=0.96 \Gamma_0$, $\delta=3.85 \Gamma_0$ and $\delta_\pi=3.99 \Gamma_0$.}
\label{dosone}
\end{figure}

We find that a gap can be opened by introducing a sublattice structure along $z$. In the halved Brillouin zone, new degeneracies arise at $|k_z|=\pi/(2a)$ due to sublattice symmetry. 
A suitable AC Stark shift can lift the degeneracy by breaking sublattice symmetry.
The ideal form of this shift can be found by considering the atom-atom interaction $\tilde{\boldsymbol{\mathcal{G}}}'$ at $|k_z|=\pi/(2a)$, which reads~\footnote{
The form of \cref{nonbravaisinteraction} can be understood intuitively: an atom in the sublattice $A$
is in the excited state $\ket{x}$ with dipole momentum $\mathbf{d}=d_0 (1,0,0)$ may radiate light
with momentum $\mathbf{k}_1=(k_x,0,\pi/(2a))$ and $ 
\mathbf{k}_2=(k_x,0,-\pi/(2a))$.
The associated electric field is $\mathbf{E}(\mathbf{r})\propto\left(\pi/(2a)\mathbf{e}_x \cos(\pi/(2a)z)-ik_x \mathbf{e}_z \sin(\pi/(2a)z)\right)\exp(ik_xx)$. (The asymmetric superposition of $e^{i\mathbf{k}_1\mathbf{r}}$ and $e^{i\mathbf{k}_2\mathbf{r}}$ is zero at the position of the emitting atom and thus decoupled.) Hence, the $x$-polarized atoms in sublattice $A$ couple to the $z$-polarized atoms in sublattice $B$ (but not to the $z$-polarization of sublattice $A$) and vice versa. The same argument can be repeated for $y$-polarization.}
\begin{equation}
  \tilde{\boldsymbol{\mathcal{G}}}'
  =\hat\id\otimes\left(\begin{array}{ccc} \tilde{\mathcal{G}}_{xx} &\tilde{\mathcal{G}}_{xy}&0\\ \tilde{\mathcal{G}}_{xy} &\tilde{\mathcal{G}}_{yy}&0\\0&0&  \tilde{\mathcal{G}}_{zz} \end{array} \right)
  	+\hat\tau_y \otimes \left(\begin{array}{ccc} 0&0&  \tilde{\mathcal{G}}_{xz} \\0&0&  \tilde{\mathcal{G}}_{yz} \\ \tilde{\mathcal{G}}_{xz} &\tilde{\mathcal{G}}_{yz}&0 \end{array}\right),\label{nonbravaisinteraction}
\end{equation}
where the sublattice structure is expressed through the Pauli matrices $\{\hat\id, \hat\tau_x,\hat\tau_y,\hat\tau_z\}$, with $\ket{A}=(1,0)$ and $\ket{B}=(0,1)$.
The $3\times3$ matrices describe the dipole--dipole interactions, where $\tilde{\mathcal{G}}_{ll'}$ are the components of the atom--atom interaction of the Bravais lattice (\emph{cf.} \cref{calcdftofg} in the appendix).

This matrix is block diagonal with two identical $3\times3$ matrices as diagonal blocks, which implies the two-fold degeneracy of each band at $|k_z|=\pi/(2a)$.

We note that to break all remaining symmetries it is not sufficient to add a perturbation of the form $H'=\tau_z\otimes\id_3$, because this leaves the symmetry between $x$ and $y$ intact.
In order to lift this symmetry as well, one can apply a magnetic field along $z$. It is then sufficient to apply the AC Stark shifts only to $\ket{\sigma_+}$ and $\ket{\pi}$, such that the perturbation reads
\begin{align}
H'&=
  \mu_0 B[\id\otimes(\ket{\sigma_+}\bra{\sigma_+}-\ket{\sigma_-}\bra{\sigma_-})]\notag\\&+ \delta
  [\hat\tau_z\otimes(\ket{\sigma_+}\bra{\sigma_+}-\ket{\pi}\bra{\pi})].
\label{perturbation}
\end{align}
Finally, we use a second AC Stark shift $\delta_\pi$ to lower the energy of the $\pi$-polarized modes, such that the gap covers the same area of energies across the whole Brillouin zone and therefore an omnidirectional band gap occurs.

To sum up, the resulting lattice is defined by lattice vectors
$\mathbf{a}_x=(a,0,0)$, $\mathbf{a}_y=(0,a,0)$ and $\mathbf{a}_z=(0,0,2a)$, where the unit cell contains two atoms A and B at positions $\mathbf{r}_A=(0,0,0)$ and $\mathbf{r}_B=(0,0,a)$ with different internal structures
\begin{align}
\omega_{+}^{A/B}&=\omega_0+\mu_B B \pm \delta\notag\\
\omega_{-}^{A/B}&=\omega_0-\mu_B B \label{levelscheme} \\
\omega_\pi^{A/B}&=\omega_0-\delta_\pi \mp \delta \notag
\end{align}
In \cref{dosone} we show that the band structure of this lattice is indeed gapped by evaluating the density of states
\begin{equation}
 \rho (\omega)=\sum_n \int_{1. BZ.} \frac{d^3k}{(2 \pi)^3} \delta (\omega-\omega_n(\mathbf{k})),
\end{equation}
where $n$ is the band index, $\omega_n(\mathbf{k})$ is the n'th band and the integral is taken over the first Brillouin zone.

Note that the ingredients introduced in the beginning of this section in principle produce a level scheme as in \cref{levelscheme}. However, all frequencies are additionally shifted by $-\delta$, which is related to the fact that the lasers presented in (iii) can only produce shifts of $-2\delta$ and $0$ at lattice sites A and B, but not shifts of $-\delta$ and $+\delta$.

\subsection{Finite size effects and defects}
In this part we briefly discuss the differences between finite and infinite lattices and the effect of defects.
Similar results have been obtained in a corresponding detailed study for the diamond lattice~\cite{antezza2013}.

\textit{Finite lattices.---}To study the effect of boundaries on the band gap, we diagonalize the full effective Hamiltonian given by Eq.~(\ref{fullheff}) for a lattice of $20\times20\times20$ atoms, and plot the resulting density of states in \cref{dosone} for both the simple cubic lattice of two-level atoms and the bipartite lattice of four-level atoms. Overall, the density of states of the finite lattice is similar to that of the infinite lattice. However, a significant difference is that some states appear in the gap, which we attribute to localized edge modes.
In finite-size lattices, the eigenstates also acquire a finite lifetime, as photons may radiate into free-space modes.
While the bulk-modes have decay rates $\Gamma_\text{bulk}\ll \Gamma_0$ that decrease with lattice size, the edge-modes are superradiant $\Gamma_\text{edge}\gg \Gamma_0$  (see \cref{appdecayfinitearray}).
For a one-dimensional chain of atoms this effect is discussed in Ref.~\cite{PhysRevX.7.031024}. As we illustrate below, this means that the infinite model provides an accurate description for impurities located deep inside the array.

\textit{Defects.---}We analyze the effect of defects by randomly removing atoms. The resulting densities of states are shown in Fig.~\ref{dosone}. Lattice defects also give rise to bound states, which contribute to the density of states in the band gap, which means that their density should be sufficiently low as to not impact the simulation.

\section{Effective impurity interactions}
\label{mediatingimpurityinteractions}

In this section we analyze the interactions mediated by band gaps in three-dimensional atomic arrays using the example when both impurity and array atoms are $\sigma_+$-polarized.
We first consider impurities placed in infinite arrays. Afterwards we generalize our results to finite arrays.

If the excitation energy of the impurities $\omega_I$ is close to the excitation energy $\omega_0$ of the array atoms, the Hamiltonian for impurities interacting with an atomic array is 
\begin{align}
H_\mathrm{tot}=H_\text{imp}+H_\text{array}+H_\text{int},
\end{align}
where the impurities are described by
\begin{align}
H_\text{imp}&=\sum_{\mathbf{r}_i} \left(\omega_I-i\frac{\Gamma_I}{2}\right)\sigma_\text{ee}^i \notag  \\ & +\frac{3 \pi c \Gamma_I }{\omega_0} \sum _{\mathbf{r}_i\neq \mathbf{r}_j}  \mathbf{d}^* \cdot \boldsymbol{\mathcal{G}}(\mathbf{r}_i-\mathbf{r}_j)\cdot \mathbf{d} \sigma_\text{eg}^i \sigma_\text{ge}^j,
\end{align}
the array Hamiltonian is given in Eq.~(\ref{fullheff}) and the interactions between impurity atoms and array atoms are
\begin{align}
H_\text{int}= \frac{3 \pi c \sqrt{\Gamma_I\Gamma_0} }{\omega_0} \sum _{\mathbf{r}_i \mathbf{R}_j}   \mathbf{d}^* \cdot \boldsymbol{\mathcal{G}}(\mathbf{r}_i-\mathbf{R}_j)\cdot  \mathbf{d}(\sigma_\text{eg}^i \hat{b}_j+\hat{b}_j^\dagger \sigma_\text{ge}^i).
\end{align}
Here, $\sigma^i$ are the spin-operators of the impurity atom at position $\mathbf{r}_i$. 
In the following we assume that every impurity atom has the same position in the respective unit cell of the array, such that all impurity atoms couple equally to the array modes.

\subsection{Infinite Array}
The bandstructure of the array is given by Eq.~(\ref{onepoldispersion}). The impurity atoms are detuned from the edge of the upper band at $\omega_c$ by $\Delta=\omega_c-\omega_I$.
If the coupling between impurity atoms and array atoms is weak ($\Gamma_I \ll \Gamma_0$), the interactions between the impurity atoms and the array can be treated under Born-Markov approximation. In this case, the full effective coupling between the impurity atoms is given by~\cite{Lehmberg1970,masson2019atomicwaveguide} 
\begin{align}
J_{ij}-i\frac{\Gamma_{ij}}{2}&=\mathcal{G}_0(\mathbf{r}_i-\mathbf{r_j})\notag \\&+ \int \frac{d^3\mathbf{k}}{(2\pi)^3}\frac{|g_{\mathbf{k}i}|^2}{\omega_I-\omega(\mathbf{k})+i0^+}e^{i \mathbf{k}\mathbf{r}_{ij}},
\label{integralforcoupling}
\end{align}
with 
\begin{align}
g_{\mathbf{k}i}=\frac{3 \pi c \sqrt{\Gamma_I\Gamma_0}}{\omega_0}\left[\sum_{\mathbf{R}} e^{-i\mathbf{k}\mathbf{R}}\ \mathbf{d}^* \cdot \boldsymbol{\mathcal{G}}(\mathbf{r}_i-\mathbf{R}) \cdot \mathbf{d} \right].
\end{align}
The first term in Eq.~(\ref{integralforcoupling}) describes the effective interactions between impurity atoms due to the exchange of free photons, while the second term takes into account modifications due to interactions between photons and array atoms. Note that the second term  describes processes where a photon emitted by the impurity excites a dressed array atom, before it is emitted again and then reabsorbed by an impurity atom. 

\begin{figure}[tb]
\includegraphics[width= \linewidth]{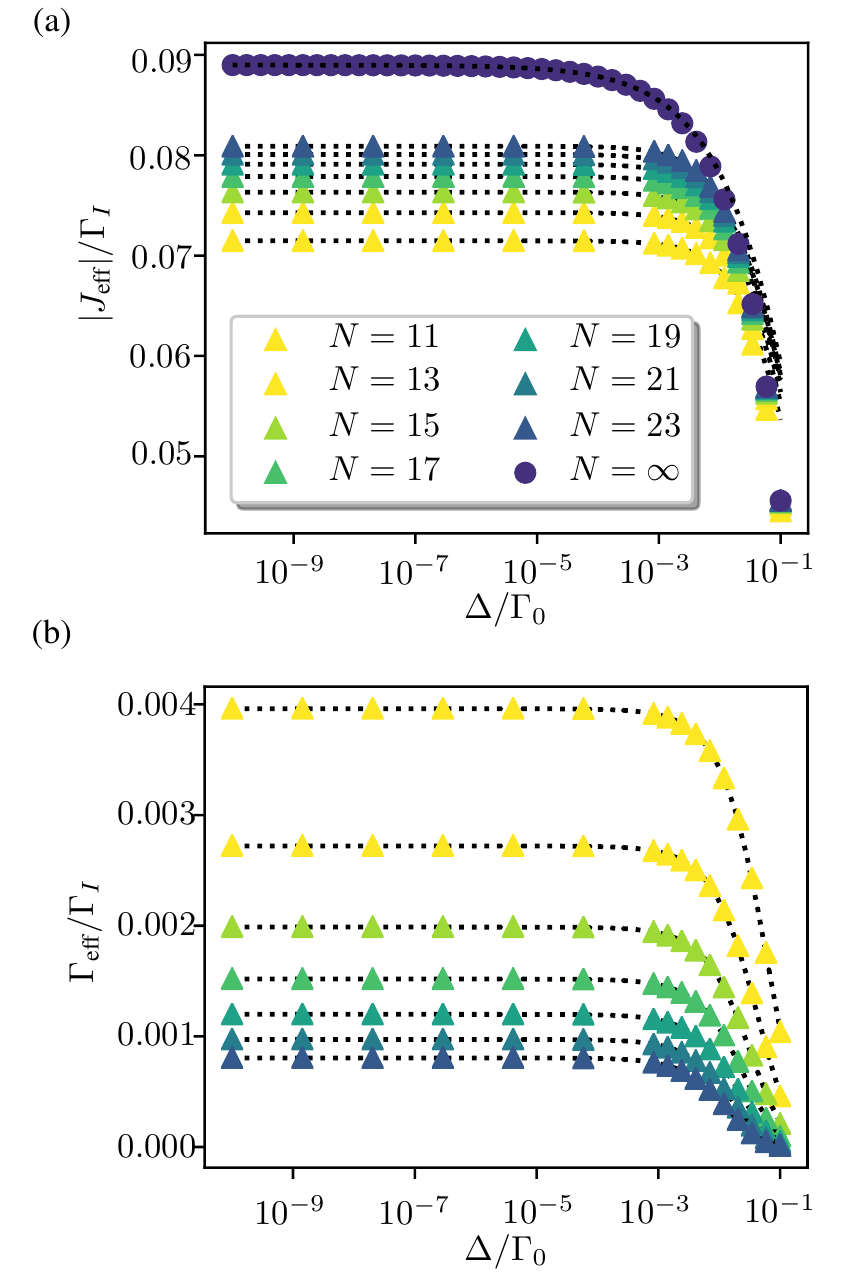}
\caption{
\textbf{Effective coupling (a) and decay (b)} of the impurities for different detunings. The infinite case (dots) is determined by numeric integration of Eq.~(\ref{integralforcoupling}). In the case of the finite lattice (crosses) the ratio of impurity and array linewidth is $\Gamma_I/\Gamma_0=10^{-3}$ and the arrays consist of $N\times N \times (N-1)$ atoms. The impurity atoms are separated by $a$ as shown in Fig.~\ref{setup}.
For the analytic approximation (gray line) we choose the following parameters to fit the numerical values (for the finite lattices the values are averaged over different system sizes):  curvature $aA^{-1/2}\approx1.28\Gamma_0^{-1/2}$ and  coupling $a^3g^2/(4\pi\sqrt{A_zA})\approx-0.089 \Gamma_I$ (infinite case, see \cref{infinite}), average curvature $a\bar{A}^{-1/2}\approx1.35\Gamma_0^{-1/2}$ and average offset $\bar{c}_1\approx2.7$ (effective coupling in finite arrays, see \cref{infinite} and \cref{cutoff}), and $a\bar{A}^{-1/2}\approx0.98\Gamma_0^{-1/2}$ and  $\bar{c}_2\approx1.38$ (effective decay in finite arrays, see \cref{eqeffectivedecay}). In general there are three free parameters for the finite cases. However, for the effective coupling in finite arrays we take the coupling $g$ from the infinite case and for the effective decay we take the offset $c_1$ from the effective coupling in finite arrays.}
\label{effectiveinteraction}
\end{figure}

The effective decay rate of an impurity atom placed in the array is given by 
\begin{align}
\Gamma_\text{eff}=\Gamma_I-2  \text{Im} \left[ \int \frac{d^3\mathbf{k}}{(2\pi)^3}\frac{|g_{\mathbf{k}i}|^2}{\omega_I-\omega(\mathbf{k})+i0^+}\right].
\label{integral}
\end{align}
In \cref{appeffectivedecay} we show that for transition frequencies in the bandgap, the imaginary part of the integral in \cref{integral} cancel the free space decay rate $\Gamma_I$ 
such that impurity atoms placed in the gap do not decay.

For impurities that are weakly detuned from the upper band edge at $\omega_c$ ($\Gamma_I \ll \Delta \ll \Gamma_0$) the effective interaction can be approximated by~\cite{de_Vega_2008}
\begin{align}
J_{ij}
&\approx \frac{a^3 g^2}{4\pi \sqrt{A_z A}} \frac{e^{-r_{ij}/\xi}}{r_{ij}},
\label{infinite}
\end{align}
where we assume the impurities to couple to a quadratic dispersion with curvature $A$ via a constant coupling $g$. The correlation length is $\xi=\sqrt{A/\Delta}$ and the effective distance we define as $r_{ij}^2=(x_i-x_j)^2+(y_i-y_j)^2+A/A_z(z_i-z_j)^2$, where the anisotropy of the interaction arises as the curvature in $k_z$ differs from the curvature in the $k_x$-$k_y$-plane. More details on the derivation of \cref{infinite} are given in \cref{appeffectivecoupling}. 

Choosing small detunings $\Delta$ one can reach the limit $r_{ij}\ll \xi$, where the coupling $J_{ij}$ is long range.    
In Fig.~\ref{effectiveinteraction} we compare the approximated coupling [Eq.~(\ref{infinite})] with the exact coupling obtained by performing the integration in Eq.~(\ref{integralforcoupling}) numerically and find good agreement.

\subsection{Finite array}
In this section we analyze how the effective interaction between impurities and the effective impurity decay change if the mediating array is finite. We consider cubic arrays with $N \times N \times N$ atoms.
For different detunings and lattice sizes the effective coupling $J_{ij}$ and the effective decay $\Gamma_{\text{eff}}$ are shown in Fig.~\ref{effectiveinteraction}.

\textit{Effective Coupling.}---The effective coupling obtained in finite arrays differs from the infinite case in the limit of $\Delta \rightarrow 0$. This can be modelled through a $N$-dependent cutoff occurring in the correlation length 
\begin{align}
  \xi_\text{fin}=\sqrt{\frac{A}{\Delta+A c_1^2/(Na)^2}}
  \label{cutoff}
\end{align} 
The cutoff arises because in a finite systems there is only a discrete set of allowed polariton moment. In particular, polaritons cannot have momentum $\mathbf{k}=0$, such that the smallest possible energy value in the upper band is $\omega_\text{min}=\omega_c+Ak_\text{min}^2$, where $k_\text{min} \propto 1/(Na)$.

\textit{Effective Decay.}---In contrast to the infinite case, we find non-vanishing effective decay rates for impurities placed in a finite array. These are caused by the fact that polaritons emitted by the impurities can decay into free space. We numerically find that the effective decay rate of an impurity placed in the middle of the atomic array scales like (Fig.~\ref{effectiveinteraction})
\begin{align}
\Gamma_\text{eff} =c_2 \Gamma_I \xi_\text{fin}^{-1}a\frac{e^{-Na/\xi_\text{fin}}}{N},
\label{eqeffectivedecay}
\end{align}
where $c_2\propto g^2/(\Gamma_0\Gamma_I)$ is a dimensionless parameter and $g$ denotes the (approximately constant) coupling of the impurity to the polariton modes.
This expression for the decay rate can be understood as the rate at which virtual polaritons that dress the impurity decay into the surrounding vacuum.
The time scale of the effective decay is then determined by the spatial distribution $\psi(\mathbf{r})=\exp(-r/\xi_\text{fin})/r$ of the polaritons and 
their average velocity $\bar{v} \propto \langle|\hat{\mathbf{p}}|\rangle \propto \xi_\text{fin}^{-1}$. The velocity of the polaritons decreases with the detuning, since for small detunings the impurity mainly couples to polariton modes with low momenta.
Importantly, \cref{eqeffectivedecay} predicts an exponential suppression of the decay rate with system size, which allows for large quality factors $Q=J_\text{eff}/\Gamma_\text{eff}$ even with moderate system sizes, such as the one studied here, which is readily achieved in experiment~\cite{Greiner2002,Tomitae1701513,PhysRevLett.94.080403}. 

\emph{Atomic motion.}---Atomic motion can severely impact the coherence in ensemble-based quantum memories~\cite{Zhao2009}. In arrays inelastic photon scattering can be suppressed by moving to the Lamb-Dicke regime, such that we neglect this effect. In order to achieve addressability and tunability, we assume that the impurity atoms are controlled with a Raman transition (see \cref{implementation}).
Since Raman transitions are slow, the motion of the atoms happens on a much faster time scale than the interactions of the atoms. In this case, the atomic motion can be eliminiated adiabatically \cite{fluctuations}. 
One then finds additional terms in the decay rate, proportional to $\eta^2\Gamma_I$, where $\eta \ll 1$ is the Lamb-Dicke parameter. Thus, provided $\eta$ is sufficiently small, one can still reach high quality factors.

\begin{figure}[tb]
\includegraphics[width=\linewidth]{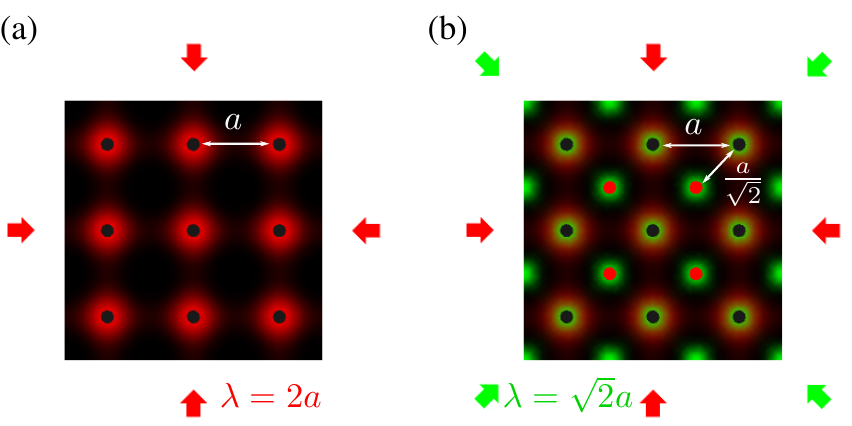}
\caption{\textbf{Configuration for loading impurities.} For clarity we show a two-dimensional version of the setup. The complete three-dimensional case is obtained by applying a standing wave along the $z$-direction. For trapping only the array atoms (black dots) one would use a lattice as shown in panel (a). The trapping potential is visualized with colors going from red (gray) to black. Red represents the nodes of the electromagnetic field, where the atoms are trapped in the case of blue-detuned trapping lasers. (b) To provide a trapping potential for the impurity atoms (red (gray) dots) as well, a second lattice (in the following called the green lattice) is needed. This lattice has nodes at the positions of the array atoms and furthermore at the desired impurity positions.} 
\label{figimplementationsetup}
\end{figure}

\begin{figure}[tb]
\centering

\includegraphics[width= \linewidth]{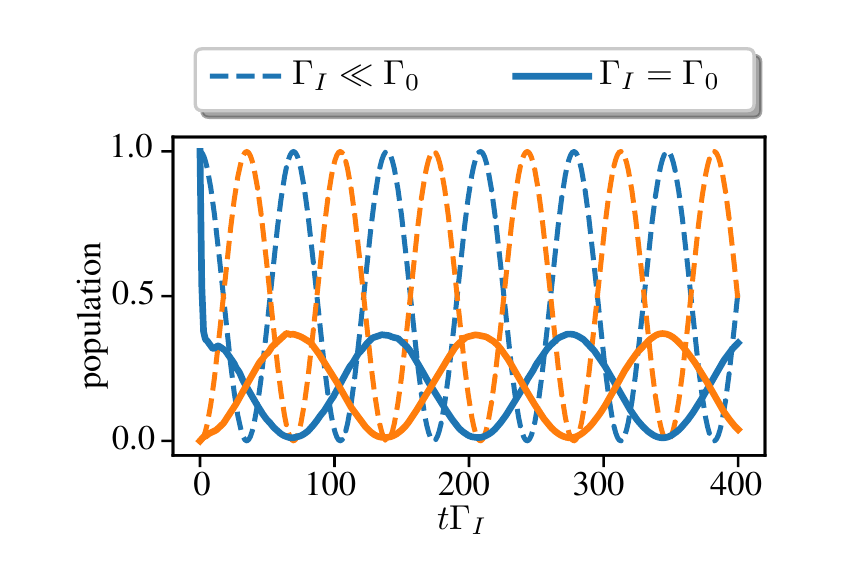}
\caption{\textbf{Rabi flopping of two impurities} in a distance of $r_{ij}=a$ placed in an array with lattice constant $a=0.4 \lambda_0$. The impurity energy is detuned from the edge of the upper band by $\Delta=0.2 \Gamma_0$. The array consists of $11\times11\times10$ atoms. The ratio of impurity linewidth and array linewidth is $\Gamma_I/{\Gamma_0}=1$ (straight lines) and $\Gamma_I/{\Gamma_0}=10^{-3}$ (dashed lines). In the strong-coupling regime the amplitude of the oscillation is smaller, as the overlap of the bound states with the initial impurity excitation is reduced. The Rabi frequency of the oscillation is $\Omega\approx0.045 \Gamma_I$ in the weak coupling regime and $\Omega\approx0.023 \Gamma_I$ for strong coupling. The effective decay rates are $\Gamma_\text{eff} \approx 3\cdot 10^{-8}\Gamma_I$ and $\Gamma_\text{eff} \approx 8\cdot 10^{-9}\Gamma_I$, respectively.}
\label{rabiflopping}
\end{figure}

\section{Experimental implementation}
\label{implementation}
In order to realize the above proposals, there are three main steps to be taken, which we discuss in detail below. 
The first is to realize an atomic array with a band gap, which, as shown above, requires an optical lattice with near unity filling but only mild subwavelength properties. 
The second step is to add impurity atoms that sit in the middle of the faces of the square lattice in $x$-$y$ direction, as shown in \cref{setup}.
Finally, in a third step, one needs some amount of control over the impurity atoms in order to excite them and read them out. 

\subsection{Three-dimensional Mott insulator}

Free-space optical lattices are a standard tool in ultracold atomic experiments~\cite{Jaksch2005}.
Mott insulators in three-dimensional optical lattices with near unity filling have been realized with both bosonic $^{87}$Rb~\cite{Greiner2002} and fermionic $^{40}$K~\cite{Schneider2008,Jordens2008}.
Since then, Mott insulators have been produced with a variety of species (\emph{e.g.}, $^{133}$Cs~\cite{Gemelke2009}, $^6$Li~\cite{Greif2016}) and are widely used for analogue quantum simulation with ultracold gases~\cite{Bloch2008,Bloch2012,Gross2017}.

To realize our proposal for an omnidirectional bandgap of one polarization, the cycling transition of $^{87}$Rb or $^{133}$Cs are ideal candidates as these species can readily be loaded into optical lattices with near-unity filling.
A photonic bandgap for both polarizations requires instead a $J=0$ to $J=1$ transition as is for example found in bosonic strontium. 
Among its bosonic isotopes, $^{84}$Sr is a suitable choice to prepare a Mott insulator state, as it can be brought into a BEC and has a suitable scattering length of around $123a_0$~\cite{Stellmer2009,Stellmer2013}.
To gap both polarizations of light, one furthermore has to apply a magnetic field in $z$-direction as well as a second laser field to produce an AC Stark shift of every other layer.
If ways are found to prepare Mott insulators with $^{86}$Sr or $^{88}$Sr (for example by mixing them~\cite{Ferrari2006}), they are also suitable candidates.

\subsection{Loading impurities}
To trap the impurities, we require a second lattice, which has nodes at the positions of the array atoms and at the intended positions of the impurity atoms as well. As illustrated in \cref{figimplementationsetup}, such a lattice can be generated using standing waves with $1/\sqrt{2}$ the original wavelength, rotated by 45 degrees relative to the standing waves of the first lattice~\cite{Lubasch2011}. In the following we call the lattice shown in \cref{figimplementationsetup} (a) the red lattice and the lattice, which is added in panel (b) the green lattice.

If the impurity atoms and the array atoms belong to the same species, one has to load them into a superposition of the two lattices discussed above.
The combined strength of the red and green lattice has to be chosen such that a Mott insulator is prepared on the red sites, and their difference, which sets the lattice depth at the impurity positions, should only allow a small density of impurities to be loaded probabilistically \footnote{The laser configuration we show in \cref{figimplementationsetup} is not the only option for realizing a lattice which has nodes at all array positions and antinodes at the impurity positions. Such a lattice can also be generated by aligning lasers with frequencies $\omega>\pi/a$ with a suitable angle between the beams.}.

In case of strontium, we have to prepare a Mott insulator of $^{84}$Sr in the red lattice, while have a small density of $^{87}$Sr as impurities in the additional green lattice sites.
One option to achieve this is to start by producing a degenerate gas mixture of bosonic $^{84}$Sr and fermionic $^{87}$Sr~\cite{Stellmer2013}.
One can then trap $^{84}$Sr with a small admixture of $^{87}$Sr in the red lattice. While the contact interactions between $^{84}$Sr--$^{84}$Sr and $^{87}$Sr--$^{87}$Sr are repulsive, with scattering lengths $123a_0$ and $96a_0$, respectively, the $^{84}$Sr--$^{87}$Sr interaction is attractive ($-57a_0$).
Thus, one can arrange that energetically, single occupancy as well as double occupancy with one $^{84}$Sr and one $^{87}$Sr are allowed, whereas double or higher occupancy the same species (or in fact any other mixtures) are disallowed.
We note here that mixtures have been loaded into the same optical lattice before~\cite{Ferrari2006,Gunter2006,Ospelkaus2006,Catani2008}.
Using state-dependent lattices, one can afterwards transfer the impurity atoms to one of the adjacent green lattice sites. Alternatively, one can excite the $^{87}$Sr atoms to the $^3$P$_0$-state, where they can be addressed specifically using a tuneout lattice~\cite{Heinz2020a}.

In fact, using tuneout lattices, one can do better. With the $^{87}$Sr atoms excited to the $^3$P$_0$ state, one can employ tweezers to deterministically control the position of the impurity atoms or sort them after loading, similar to how arrays of Rydberg atoms have been assembled~\cite{Endres2016}.
This is a particularly exciting avenue, as it would give access to fully controllable lattices of impurities within the host medium.

\subsection{Impurity control}

To control the coupling strength as well as the transition frequency of the impurity atoms, we suggest to use a $\Lambda$-scheme~\cite{Gonzalez-Tudela2015,Hu2017,Urvoy2019}.
An ideal two-level Raman scheme should involve the cycling transition in Rb ($\ket{g_1}=\ket{5S_{1/2},F=2,m_F=2} \rightarrow \ket{e}=\ket{5P_{1/2},F=3,m_F=3}$). As discussed in Ref.~\cite{Porras2008}, this can be achieved using a two-photon transition from a second hyperfine ground state ($\ket{g_2}=\ket{5S_{1/2},F=1,m_F=1}$).
Examples for the intermediate state of this two-photon transition are $\ket{5S_{1/2},F=2,m_F=2}$ or a $6\text{P}_{3/2}$ state. The former requires a microwave laser to couple the two hyperfine ground states and one to drive the cycling transition, whereas the latter option requires a laser to drive the $\ket{g_2}$-$6\text{P}_{3/2}$-transition and one to drive the $\ket{e}$-$6\text{P}_{3/2}$-transition.
In the case of \textsuperscript{87}Sr one can use a $6^1\text{P}_{1}$ state to couple the cycling transition ($\ket{g_1}=\ket{5^1S_{0},F=9/2,m_F=9/2} \rightarrow \ket{e}=\ket{5^1P_{1},F=11/2,m_F=11/2}$ ) to a second hyperfine ground state (e.g. $\ket{g_2}=\ket{5^1S_{0},F=9/2,m_F=7/2}$).

Either way, one must ensure that the Raman lasers do not affect the array atoms. For the optical laser, a possibility is to use the lasers generating the optical lattice of the array atoms, which have nodes at the positions of the array atoms, but not the impurity atoms, at the expense of limiting the allowed lattice spacings $a$. This is not possible for the microwave tone, which instead has to be far detuned, such that its effect on the array atoms is negligible. More details are given in \cref{applambdawithsr}.

Finally we note that as shown in Fig.~\ref{rabiflopping}, our proposal is not restricted to weak coupling, such that the implementation of a $\Lambda$-scheme is not mandatory. In the strong coupling regime, the transition frequency could be tuned using AC Stark shifts generated by the array lasers.

To excite impurities, we suggest to use two-photon transitions. First, we consider the case where the array atoms and the impurity atoms belong to the same species, using the example of $^{87}$Rb. We assume that all atoms are initially prepared in the same ground state (e.g. $\ket{g_1}=\ket{5S_{1/2},F=2,m_F=2}$). Using a $6\text{P}_{3/2}$ state one can then engineer transitions from $\ket{g_1}$ to $\ket{g_2}=\ket{5S_{1/2},F=1,m_F=1}$. Here, the lasers driving the transition from $\ket{g_1}$ to $6\text{P}_{3/2}$ have to be aligned such that the resulting electric field has nodes at all array positions, such that only impurity atoms are excited. This is possible for lattice constants $a>210 $ nm. To reach smaller lattice constants one should choose an intermediate state with a higher transition frequency. An electric field with nodes at all array positions can also be used to ensure that the frequencies of the $\ket{g_1}$-$6\text{P}_{3/2}$-transitions of the impurity atoms and array atoms differ. One can then shape the excitation laser, which does no longer effect array atoms, to capture single sites such that selected impurity atoms can be excited.

If the array atoms and the impurity atoms belong to different species ($^{84}$Sr and $^{87}$Sr) their transition frequencies differ naturally, such that single-site addressing is more straightforward. Considering strontium we assume that all impurity atoms are prepared in the state $\ket{g_1}=\ket{5^1S_{0},F=9/2,m_F=9/2}$. One can then transfer individual impurity atoms into $\ket{g_2}=\ket{5^1S_{0},F=9/2,m_F=7/2}$ for example via a $6^1\text{P}_{1}$-state.

The simplest way to read out the impurities is to drop the array atoms and then image the remaining impurity atoms. A less invasive technique would first transfer the excited impurities to a different level with a cycling transition, such that they can be imaged without losing the array atoms.

\section{Simulation of spin systems with long-range interactions}
\label{quantumsimulation}

Our proposal can be used to engineer effective spin Hamiltonians of the form $H=\sum_{ij} J_{ij}\sigma_i^+\sigma_j^-$, where the effective interactions $J_{ij}$ are given in \cref{infinite}.
Using $\Lambda$ and four-level systems the XXZ-model or the transverse Ising model for spin-1/2 can also be realized \cite{Douglas_2015,Gonzalez-Tudela2015}.
For short correlation lengths $\xi<a$, this includes nearest-neighbour interacting spin models.

As we argue in this section, this platform is also capable of simulating long-range interacting spin models if the system parameters are tuned properly. 
The important length scales are the correlation length $\xi$, defined in \cref{cutoff}, the size of the array, which is taken to be a cube of side length $Na$, and the size of the embedded impurity system, which is (say) a square of size $L\times L$.
%

Long-range interactions can be achieved when the correlation length $\xi$ is much larger than the system $\xi\gg L$, as in this regime, the effect of the exponential envelope becomes negligible.
For example, if we take a reasonably sized system with $L=8a$ and a correlation length $\xi=15a$, then the correction to the $1/r$ interaction is at most $\exp(-\sqrt{2} \ 8/15)\approx 1/2$ for the pair of impurities furthest away from each other.

To suppress excitation loss due to decay into free-space photons, the correlation length must at the same time be much smaller than the array size $\xi\ll Na$, or specifically the smallest distance between an impurity atom and the edge of the array. 
If we assume the 8-by-8 grid of impurities to be embedded in an array of size $N=50$, which could be achieved with slight improvements over the state of the art \cite{Fukuhara2009}, the smallest distance to the edge is $20.5 a$, leading to a ratio of the fastest decay rate to the fastest interaction rate of 571, which implies that the simulation could probe many Rabi cycles.

Finally, in a usable simulation, the interaction time scale $\tau_\text{int}\propto 1/\Gamma_I$ should not become exceedingly long, as there is an upper limit to how long atoms can stay trapped and their hyperfine states coherent, which we could optimistically set at $1s$.
Physically, to simulate long-range interactions, we require that the propagation time of polaritons from one impurity atom to the next is negligible ($\tau_\text{int}\gg L/\bar{v}_\text{pol}$) and at the same time the detuning has to be small, which in turn reduces the group velocity ($\bar{v}_\text{pol} \propto L^{-1}$) of the polaritons as more slow polariton modes near the band edge are admixed.
Combining these two requirements lead to a scaling  $\tau_\text{int}=\mathcal O(L^{2})$. This corresponds to the assumption $\Gamma_I\ll \Delta$, where $\Delta=A \xi^{-2}=\mathcal{O}(L^{-2})$ in the limit of long-range interactions.
In our example above, we find $J=0.083 \Gamma_I$, for atoms in a distance $r=a$ with a correlation length $\xi=15 a$.

\section{Conclusion}
\label{conclusion}
We have proposed to use three-dimensional atomic arrays with bandgaps as nanophotonic metamaterials.
We have shown that inserting impurity atoms whose transition frequencies lie in the band gap
these setups can be used to engineer effective interactions with exponentially suppressed decay rates. This allows the implementation of effective spin Hamiltonians. 
While we concentrated on the simulation of spin 1/2 systems using two-level atoms, we expect that tuning interactions between three- and four-level impurity atoms may allow one to implement higher spin models. 
Our proposal uses only one specific feature of the three-dimensional bandstructure of this metamaterial, namely the quadratic dispersion near the band edge. It is known that many novel, non-Markovian effects occur in the presence of three-dimensional structured reservoirs~\cite{Gonzalez-Tudela2018}, which is another exciting direction to take this platform.
While in this paper we focused on simple cubic atomic arrays for engineering bandgaps, we expect that our approach can be extended to different geometries, which should be explored in future research.

Our proposal allows the implementation of unitary dipole--dipole interactions for systems in one-, two-, and even three-dimensional systems and is compatible with current ultracold atomic quantum simulators.
Tunable spin--spin interactions enable the exploration of exciting new physics, including novel quantum spin phases~\cite{DallaTorre2006,Yao2012,Yao2013,Manmana2013,Gong2016,Baier2016}, the competition between short- and long-range interactions~\cite{Landig2016}, or frustration~\cite{Sandvik2010}. Furthermore, the possibility to implement coulomb-like interactions between localized states allows the study of electron glasses, which are known to posses phenomena as slow relaxation and aging~\cite{Amir2009}. 

In the future one might envision integrating light-mediated interactions with standard Bose- or Fermi-Hubbard quantum simulators to access a rich family of Hamiltonians.
Atomic metamaterials with bandgaps such as the one studied here may also be used to shield or capture radiation in a very specific frequency range and thus may find uses beyond quantum simulation.

\begin{acknowledgements}
We thank Monika Aidelsburger, Robert Bettles, Ignacio Cirac, Thomas Kohlert, Annie Park, Jun Rui, Sebastian Scherg and Konrad Viebahn for insightful discussions, and Cosimo Rusconi and Alejandro Gonzalez-Tudela for comments on the manuscript.
DM acknowledges funding from ERC Advanced Grant QUENOCOBA under the EU Horizon 2020 program (Grant Agreement No. 742102).
\end{acknowledgements}

\appendix 

\section{Green's function}
\label{greensfunction}
Here, we present details about the Greens' function used in  in Eq.~(\ref{fullheff}) and its Fourier transform.
In free space the Green's tensor is given by the dyadic Green's function, evaluated at the atomic transition frequency $\omega_0$~\cite{Lehmberg1970,Lehmberg1970a} 
\begin{align}
  \mathcal{G}_{ll'}(\mathbf{r})&=-\frac{e^{ik_0r}}{4 \pi k_0^2r^3} \left[(k_0^2r^2+ik_0r-1)\delta_{ll'}\right.\notag \\ &+\left. (3-3ik_0r-k_0^2r^2)\frac{r_l r_{l'}}{r^2}\right]+\frac{\delta_{ll'}\delta^{(3)}(\mathbf{r})}{3k_0^2},
\end{align}
where $k_0=\omega_0/c$ is the resonant wavevector and $l,l'=x,y,z$ label the spatial directions. 
To calculate the atom--atom-interactions in Eq.~(\ref{main}), we use the approximation 
\begin{align}
\sum_{\mathbf{R} \neq 0} \mathcal{G}_{ll'}(\mathbf{R}) e^{i\mathbf{k}\mathbf{R}} \approx e^{k_0^2a_\text{ho}^2/2} \left[  \frac{1}{V_\text{L}} \sum _{\mathbf{G}}  g_{ll'}'(\mathbf{k}+\mathbf{G})-\mathcal{G}_{ll'}'(0)\right].
\label{calcdftofg}
\end{align}
Here, the quantum fluctuations $a_\text{ho}$ of the atomic positions were introduced to avoid divergencies~\cite{PhysRevA.96.063801,fluctuations,antezza} and $V_L$ is the volume of the unit cell. 
The Fourier transform of the regularized the Green's function is given by
\begin{align}
g_{ll'}'(\mathbf{k})=\frac{1}{k_0^2} \frac{k_0^2 \delta_{ll'} -k_l k_{l'}}{k_0^2-k^2}e^{-k^2a_\text{ho}^2/2} 
\label{eqfourietrafofg}
\end{align} 
and the regularized Green's function at $\mathbf{r}=0$ is \\ $\mathcal{G}_{ll'}'(0)=\delta_{ll'} \mathcal{G}'(0)$, with
\begin{align}
\mathcal{G}'(0)=\frac{k_0}{6\pi}\left[\frac{\text{Erfi}(k_0a_0/\sqrt{2})-i}{e^{(k_0a_0)^2/2}}-\frac{-1/2+(k_0a_0)^2}{\sqrt{\pi/2}(k_0a_0)^3}\right],
\end{align}
where $\text{Erfi}(x)=2\pi^{-1/2}\int_0^x dy \exp(y^2)$ is the imaginary error function.

\section{Decay rate of array modes in finite systems}
\label{appdecayfinitearray}
\begin{figure*}[tb]
\includegraphics[width=\linewidth]{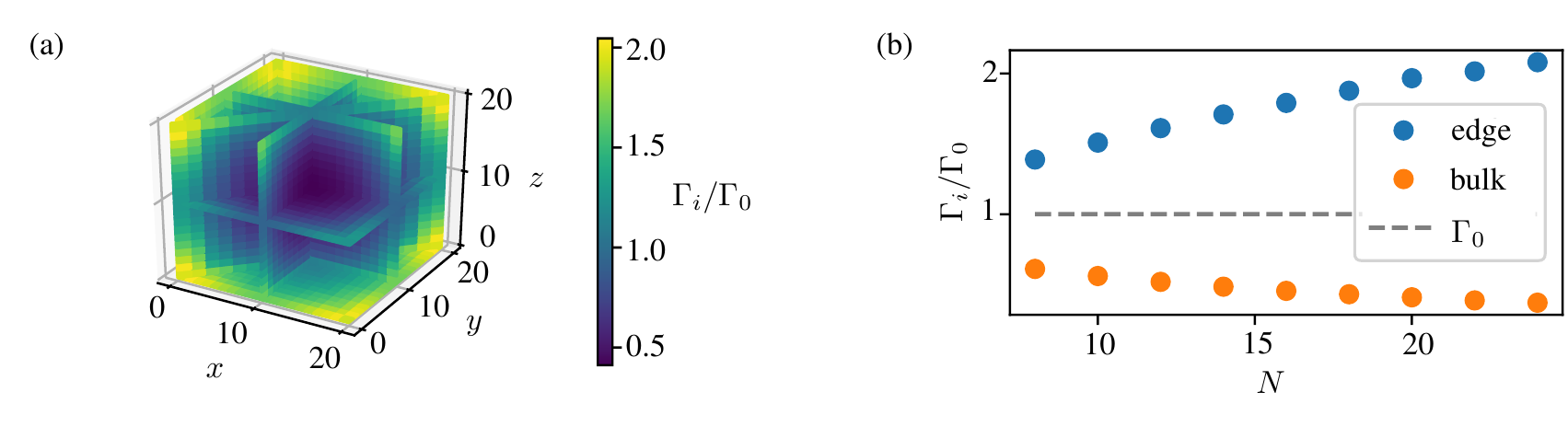}
\caption{\textbf{Average decay rate} for different atomic positions (a) and different system sizes (b). a) One finds that the average decay rate at the edges is larger than $\Gamma_0$ while the average decay in the bulk is suppressed. b) The average decay of an atom sitting in the corner increases with $N$, while the average decay of an atom in the middel decreases.}
\label{averagedecay}
\end{figure*}
Here, we discuss the finite lifetimes of the array modes of finite systems. To illustrate that the strongly decaying modes are mainly localized at the edge of the array, we calculate the average decay rate for each atomic position. Assuming that the eigenmodes of the finite array are $\ket{\xi}$ with decay rates $\Gamma_\xi$, the average decay rate $\bar{\Gamma}_i$ of the atom at position $\mathbf{R}_i$ is given by
\begin{align}
\bar{\Gamma}_i=\sum_\xi \Gamma_\xi|\braket{\mathbf{R}_i|\xi}|^2.
\end{align}
In Fig.~\ref{averagedecay} we show the average decay rate. The strongest average decay one finds at the corners of the array. In the middle of the array the average decay is the smallest. While the atoms in the middle are surrounded by atoms that protect them from decaying, the atoms at the edges can radiate into free space. For different system sizes we compare the average decay rate of the atoms at the corner with that of the atoms in the middle in Fig.~\ref{averagedecay}. While the average decay rate of atoms in the bulk decreases with increasing $N$, the average decay rate of the atoms at the edge increases. Similar effects have been observed in~\cite{PhysRevX.7.031024}.


\section{Effective impurity decay in infinite systems}
\label{appeffectivedecay}
Here, we show that impurity atoms which are placed in infinite arrays do not decay if their transition frequency lies in the bandgap. Up to second order, the decay rate is described by \cref{integral}. Using the Poisson summation formula we write the coupling $g_{\mathbf{k}i}$ and the dispersion $\omega(\mathbf{k})$ as
\begin{align}
g_{\mathbf{k}i}=\frac{3 \pi c \sqrt{\Gamma_I\Gamma_0}}{\omega_0 \sqrt{N^3}V_\text{L}} \sum_\mathbf{G}\mathbf{d}^* \cdot \mathbf{g}' (\mathbf{k}-\mathbf{G})  \cdot \mathbf{d} e^{-i\mathbf{G} \cdot \mathbf{r}_i}   
\end{align}
and
\begin{align}
\omega(\mathbf{k})=\frac{3 \pi c \Gamma_0}{\omega_0 V_\text{L}} \sum_\mathbf{G}\mathbf{d}^* \cdot \mathbf{g}'(\mathbf{k}-\mathbf{G})  \cdot \mathbf{d},  
\end{align}
where the Fourier transform of the Greens function $\mathbf{g}'(\mathbf{k})$ is given by \cref{eqfourietrafofg}. As the integrand in \cref{integral} is real, only the poles of the integrand are relevant for the decay rate. If the energy of the impurity atoms is placed in the band gap, there is only one pole at $k=k_0$. Near $k_0$, the $\mathbf{G}=0$-term in $g_{\mathbf{k}i}$ and $\omega(\mathbf{k})$ diverges. At this point, for $k \in I_\epsilon=[k_0-\epsilon,k_0+\epsilon]$ the integral in \cref{integral} can be approximated by
\begin{align}
V &\int_{k \in I_\epsilon} \frac{d^3k}{(2\pi)^3}\frac{|g_{\mathbf{k}i}|^2}{\omega_I-\omega(k)+i0^+}\notag \\&\approx \frac{3 \pi \Gamma_I}{2k_0^3} \int_{k \in I_\epsilon} \frac{d^3k}{(2\pi)^3}\frac{|\mathbf{d}^* \cdot \mathbf{g}'(\mathbf{k})  \cdot \mathbf{d}|^2}{-\mathbf{d}^* \cdot \mathbf{g}'(\mathbf{k})  \cdot \mathbf{d}} \notag \\ &=-\frac{3 \pi \Gamma_I}{2k_0^3}\int_{k \in I_\epsilon} \frac{d^3k}{(2\pi)^3} \mathbf{d}^* \cdot \mathbf{g}'(\mathbf{k})  \cdot \mathbf{d}.
\end{align}
Using the Sokhotski-Plemelj theorem one obtains
\begin{align}
&\int_{k \in I_\epsilon} \frac{d^3k}{(2\pi)^3} \mathbf{d}^* \cdot \mathbf{g}'(\mathbf{k})  \cdot \mathbf{d} \notag \\ &= \frac{1}{(2\pi)^2}\frac{8}{3} \int_{k_0-\epsilon}^{k_0+\epsilon} \frac{k^4}{k_0^2-k^2+i0^+} \notag \\ &=-i\pi\frac{1}{(2\pi)^2}\frac{4}{3}k_0^3,
\end{align}
such that in total $\Gamma_\text{eff}=0$.

\section{Effective impurity interactions}
\label{appeffectivecoupling}
To obtain an analytic understanding of the effective coupling between impurity atoms, we use the effective mass approximation. The approximated dispersion takes the form $\omega(\mathbf{k})=A(k_x^2+k_y^2)+A_zk_z^2$, where we have included the fact that the curvature along $k_z$ differs from the curvature along $k_x$ and $k_y$ (see \cref{setup}). 
Furthermore we assume that the coupling $g_\mathbf{k}$ is constant such that the effective interaction takes the form
\begin{align}
J_{ij}&=a^3 g^2 \int \frac{d^3k}{(2\pi)^3}\frac{1}{\Delta+A(k_x^2+k_y^2+A_z/Ak_z^2)}e^{i \mathbf{k}\mathbf{r}_{ij}}.
\end{align}

\begin{figure*}[tb]
\includegraphics[width= \linewidth]{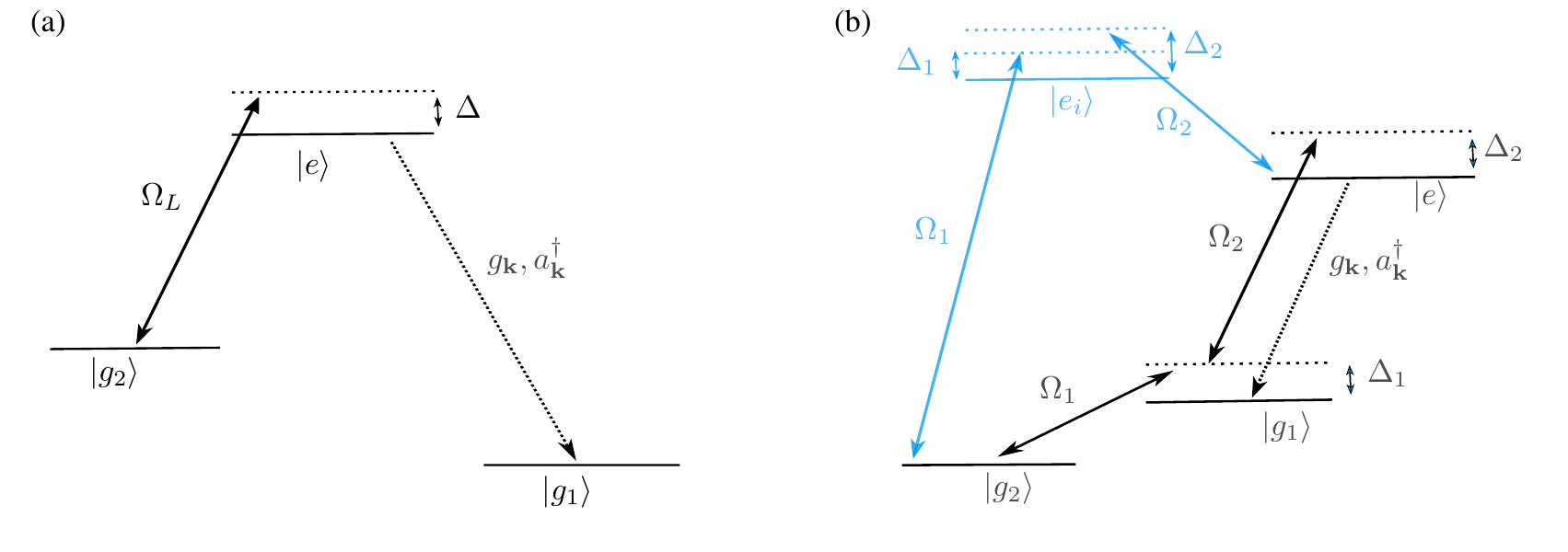}
\caption{(a) A lambda scheme to tune the transition frequency and coupling strength of impurity atoms.
(b) A two-photon lambda scheme allows one to use the cycling transition~\cite{Porras2008}. It is equivalent to a $\Lambda$-scheme with $\Omega_L=\Omega_1\Omega_2/\Delta_1$ (if $\Delta_2 \ll \Delta_1$). One can also couple $\ket{g_2}$ and $\ket{e}$ via an additional state $\ket{e_i}$ (blue scheme). }
\label{lambdascheme}
\end{figure*}
Substituting $\sqrt{A_z/A}k_z$ with $k_z$ and $\sqrt{A/A_z}z$ with $z$ one finds
\begin{align}
J_{ij}&=a^3 g^2 \sqrt{\frac{A}{A_z}}\int \frac{d^3k}{(2\pi)^3}\frac{1}{\Delta+Ak^2}e^{i \mathbf{k}\mathbf{r}_{ij}}\notag\\&=\frac{a^3 g^2}{2\pi^2} \sqrt{\frac{A}{A_z}} \int_0^{k_c} dk\frac{k}{\Delta+Ak^2}\frac{\sin(kr_{ij})}{r_{ij}}
\notag\\ &=\frac{a^3 g^2}{4\pi^2\sqrt{A_zA}} \int_0^{q'_c} dq'\frac{q'}{1+q'^2}\frac{e^{iq'r_{ij}/\xi}-e^{-iq'r_{ij}/\xi} }{ir_{ij}}\notag\\
&\approx \frac{a^3 g^2}{4\pi \sqrt{A_zA}} \frac{e^{-r_{ij}/\xi}}{r_{ij}},
\label{supinfinite}
\end{align}
where we introduced $q=\xi k$ with the correlation length $\xi=\sqrt{A/\Delta}$ and furthermore assumed  $\xi \rightarrow \infty$, which is a valid assumption for small detunings $\Delta$.

\section{$\Lambda$-schemes with $^{87}$Rb and $^{87}$Sr}
\label{applambdawithsr}

As mentioned before, a $\Lambda$-scheme can be used to tune the transition frequency and the coupling strength of the impurity atoms. In particular, a $\Lambda$-scheme as shown in \cref{lambdascheme} gives rise to an effective coupling of the form~\cite{Gonzalez-Tudela2015}
\begin{align} 
H_\text{int}=-\sum_k g_k \frac{\Omega_L}{2\Delta} (a_k^\dagger  \ket{g_1}\bra{g_2} +h.c.).
\label{finalh}
\end{align}
Here, we discuss how this can be implemented with $^{87}$Rb and $^{87}$Sr.

The cycling transition of $^{87}$Rb with $\ket{g_1}=\ket{5S_{1/2},F=2,m_F=2}$ and $\ket{e}=\ket{5P_{1/2},F=3,m_F=3}$ can be mapped to a $\Lambda$-scheme using a level configuration as presented in \cref{lambdascheme}, where the second hyperfine ground state could be choosen as $\ket{g_2}=\ket{S_{1/2},F=1,m_F=1}$~\cite{Porras2008}.
To achieve that the resonantly emitted photons have frequencies lying in the bandgap the second laser has to satisfy
\begin{align}
\Delta_1+\omega_{L,2} \approx \omega_e^{array}-\omega_g^{array},
\label{condition}
\end{align}
which implies that the second laser is near-resonant to the cycling transition of the array atoms. To avoid any couplings between this laser and the array it is thus reasonable to use one of the lasers generating the optical lattice of the array atoms. In this case, the detuning $\Delta_1$ is determined by the detuning of the trapping laser. Since $\Delta_1$ has to be of the order of GHz, the detuning of the trapping laser is small and the lattice spacing $a$ is close to $\lambda_0/2$. 

To obtain setups with smaller lattice spacings one could engineer effective couplings between $\ket{g_2}$ and $\ket{e}$ using $\ket{e_i}=\ket{6P_{1/2},F=3,m_F=2}$ as an intermediate state (see also \cref{lambdascheme}). The lasers driving the $\ket{g_2}$-$\ket{e_i}$-transition can be aligned such that the resulting electric field has nodes at all array positions.

Another possibility is to use $^{84}$Sr for the array and $^{87}$Sr for the impurity atoms.
$^{87}$Sr has a nuclear spin of $I=9/2$, which gives rise to hyperfine structure splitting. One possible choice for implementing a lambda scheme is 
$\ket{g_1}=\ket{5^1S_{0},F=9/2,m_F=9/2}$, $\ket{g_2}=\ket{5^1S_{0},F=9/2,m_F=7/2}$, $ \ket{e}=\ket{5^1P_{1},F=11/2,m_F=11/2}$ and $\ket{e_i}=\ket{6^1P_{1},F=11/2,m_F=9/2}$. Since the transition frequencies of $^{84}$Sr and $^{87}$Sr differ, the raman lasers are far detuned from the transitions of the array atoms, such that the couplings between these lasers and the array atoms are negligible.

Some of the lasers discussed above do also couple to the array atoms. Here, we discuss their effect using the example of the microwave laser, which is used to couple two hyperfine ground states in $^{87}$Rb. Up to first order, the perturbed ground state $\ket{g_1}$ takes the form
\begin{align}
\ket{g_1}=\ket{g_1}-\Omega_1/\Delta_1\ket{g_2}
\end{align}
The probability that an atom initially in $\ket{g_1}$ is excited to $\ket{g_2}$ after switching on $\Omega_1$ is thus $p_{g_2}=(\Omega_1/\Delta_1)^2$.
Declaring the atoms in $\ket{g_2}$ as additional defects, this effect can be neglected if the defect density due to the microwave laser $p_{g_2}=N_s/N_\mathrm{tot}$ is much smaller than the defect density due to finite filling. For example, if $N_\text{def}/N_\mathrm{tot} \gtrsim 0.1$~\cite{Rui2020}, we require $(\Omega_1/\Delta_1)^2\ll 0.1$. The laser driving the cycling transition has to be much stronger than the microwave tone to achieve sufficiently strong effective Raman transition rates $\Omega_L=\Omega_1 \Omega_2/\Delta_1$.


\bibliography{references,library}

\begin{thebibliography}{74}%
\makeatletter
\providecommand \@ifxundefined [1]{%
 \@ifx{#1\undefined}
}%
\providecommand \@ifnum [1]{%
 \ifnum #1\expandafter \@firstoftwo
 \else \expandafter \@secondoftwo
 \fi
}%
\providecommand \@ifx [1]{%
 \ifx #1\expandafter \@firstoftwo
 \else \expandafter \@secondoftwo
 \fi
}%
\providecommand \natexlab [1]{#1}%
\providecommand \enquote  [1]{``#1''}%
\providecommand \bibnamefont  [1]{#1}%
\providecommand \bibfnamefont [1]{#1}%
\providecommand \citenamefont [1]{#1}%
\providecommand \href@noop [0]{\@secondoftwo}%
\providecommand \href [0]{\begingroup \@sanitize@url \@href}%
\providecommand \@href[1]{\@@startlink{#1}\@@href}%
\providecommand \@@href[1]{\endgroup#1\@@endlink}%
\providecommand \@sanitize@url [0]{\catcode `\\12\catcode `\$12\catcode
  `\&12\catcode `\#12\catcode `\^12\catcode `\_12\catcode `\%12\relax}%
\providecommand \@@startlink[1]{}%
\providecommand \@@endlink[0]{}%
\providecommand \url  [0]{\begingroup\@sanitize@url \@url }%
\providecommand \@url [1]{\endgroup\@href {#1}{\urlprefix }}%
\providecommand \urlprefix  [0]{URL }%
\providecommand \Eprint [0]{\href }%
\providecommand \doibase [0]{https://doi.org/}%
\providecommand \selectlanguage [0]{\@gobble}%
\providecommand \bibinfo  [0]{\@secondoftwo}%
\providecommand \bibfield  [0]{\@secondoftwo}%
\providecommand \translation [1]{[#1]}%
\providecommand \BibitemOpen [0]{}%
\providecommand \bibitemStop [0]{}%
\providecommand \bibitemNoStop [0]{.\EOS\space}%
\providecommand \EOS [0]{\spacefactor3000\relax}%
\providecommand \BibitemShut  [1]{\csname bibitem#1\endcsname}%
\let\auto@bib@innerbib\@empty
\bibitem [{\citenamefont {Noh}\ and\ \citenamefont
  {Angelakis}(2017)}]{Noh2017}%
  \BibitemOpen
  \bibfield  {author} {\bibinfo {author} {\bibfnamefont {C.}~\bibnamefont
  {Noh}}\ and\ \bibinfo {author} {\bibfnamefont {D.~G.}\ \bibnamefont
  {Angelakis}},\ }\bibfield  {title} {\bibinfo {title} {{Quantum simulations
  and many-body physics with light}},\ }\href
  {https://doi.org/10.1088/0034-4885/80/1/016401} {\bibfield  {journal}
  {\bibinfo  {journal} {Reports on Progress in Physics}\ }\textbf {\bibinfo
  {volume} {80}},\ \bibinfo {pages} {016401} (\bibinfo {year}
  {2017})}\BibitemShut {NoStop}%
\bibitem [{\citenamefont {González-Tudela}\ \emph {et~al.}(2015)\citenamefont
  {González-Tudela}, \citenamefont {Hung}, \citenamefont {Chang},
  \citenamefont {Cirac},\ and\ \citenamefont {Kimble}}]{Gonzalez-Tudela2015}%
  \BibitemOpen
  \bibfield  {author} {\bibinfo {author} {\bibfnamefont {A.}~\bibnamefont
  {González-Tudela}}, \bibinfo {author} {\bibfnamefont {C.-L.}\ \bibnamefont
  {Hung}}, \bibinfo {author} {\bibfnamefont {D.~E.}\ \bibnamefont {Chang}},
  \bibinfo {author} {\bibfnamefont {J.~I.}\ \bibnamefont {Cirac}},\ and\
  \bibinfo {author} {\bibfnamefont {H.~J.}\ \bibnamefont {Kimble}},\ }\bibfield
   {title} {\bibinfo {title} {Subwavelength vacuum lattices and atom-atom
  interactions in two-dimensional photonic crystals},\ }\href
  {https://doi.org/10.1038/nphoton.2015.54} {\bibfield  {journal} {\bibinfo
  {journal} {Nature Photonics}\ }\textbf {\bibinfo {volume} {9}},\ \bibinfo
  {pages} {320} (\bibinfo {year} {2015})}\BibitemShut {NoStop}%
\bibitem [{\citenamefont {Douglas}\ \emph {et~al.}(2015)\citenamefont
  {Douglas}, \citenamefont {Habibian}, \citenamefont {Hung}, \citenamefont
  {Gorshkov}, \citenamefont {Kimble},\ and\ \citenamefont
  {Chang}}]{Douglas_2015}%
  \BibitemOpen
  \bibfield  {author} {\bibinfo {author} {\bibfnamefont {J.~S.}\ \bibnamefont
  {Douglas}}, \bibinfo {author} {\bibfnamefont {H.}~\bibnamefont {Habibian}},
  \bibinfo {author} {\bibfnamefont {C.-L.}\ \bibnamefont {Hung}}, \bibinfo
  {author} {\bibfnamefont {A.~V.}\ \bibnamefont {Gorshkov}}, \bibinfo {author}
  {\bibfnamefont {H.~J.}\ \bibnamefont {Kimble}},\ and\ \bibinfo {author}
  {\bibfnamefont {D.~E.}\ \bibnamefont {Chang}},\ }\bibfield  {title} {\bibinfo
  {title} {Quantum many-body models with cold atoms coupled to photonic
  crystals},\ }\href {https://doi.org/10.1038/nphoton.2015.57} {\bibfield
  {journal} {\bibinfo  {journal} {Nature Photonics}\ }\textbf {\bibinfo
  {volume} {9}},\ \bibinfo {pages} {326–331} (\bibinfo {year}
  {2015})}\BibitemShut {NoStop}%
\bibitem [{\citenamefont {Hood}\ \emph {et~al.}(2016)\citenamefont {Hood},
  \citenamefont {Goban}, \citenamefont {Asenjo-Garcia}, \citenamefont {Lu},
  \citenamefont {Yu}, \citenamefont {Chang},\ and\ \citenamefont
  {Kimble}}]{Hood10507}%
  \BibitemOpen
  \bibfield  {author} {\bibinfo {author} {\bibfnamefont {J.~D.}\ \bibnamefont
  {Hood}}, \bibinfo {author} {\bibfnamefont {A.}~\bibnamefont {Goban}},
  \bibinfo {author} {\bibfnamefont {A.}~\bibnamefont {Asenjo-Garcia}}, \bibinfo
  {author} {\bibfnamefont {M.}~\bibnamefont {Lu}}, \bibinfo {author}
  {\bibfnamefont {S.-P.}\ \bibnamefont {Yu}}, \bibinfo {author} {\bibfnamefont
  {D.~E.}\ \bibnamefont {Chang}},\ and\ \bibinfo {author} {\bibfnamefont
  {H.~J.}\ \bibnamefont {Kimble}},\ }\bibfield  {title} {\bibinfo {title}
  {Atom{\textendash}atom interactions around the band edge of a photonic
  crystal waveguide},\ }\href {https://doi.org/10.1073/pnas.1603788113}
  {\bibfield  {journal} {\bibinfo  {journal} {Proceedings of the National
  Academy of Sciences}\ }\textbf {\bibinfo {volume} {113}},\ \bibinfo {pages}
  {10507} (\bibinfo {year} {2016})}\BibitemShut {NoStop}%
\bibitem [{\citenamefont {Liu}\ and\ \citenamefont {Houck}(2017)}]{liu2017}%
  \BibitemOpen
  \bibfield  {author} {\bibinfo {author} {\bibfnamefont {Y.}~\bibnamefont
  {Liu}}\ and\ \bibinfo {author} {\bibfnamefont {A.}~\bibnamefont {Houck}},\
  }\bibfield  {title} {\bibinfo {title} {Quantum electrodynamics near a
  photonic bandgap},\ }\href {https://doi.org/10.1038/nphys3834} {\bibfield
  {journal} {\bibinfo  {journal} {Nature Physics}\ }\textbf {\bibinfo {volume}
  {13}},\ \bibinfo {pages} {48} (\bibinfo {year} {2017})}\BibitemShut {NoStop}%
\bibitem [{\citenamefont {Chang}\ \emph {et~al.}(2018)\citenamefont {Chang},
  \citenamefont {Douglas}, \citenamefont {Gonz{\'{a}}lez-Tudela}, \citenamefont
  {Hung},\ and\ \citenamefont {Kimble}}]{Chang2018}%
  \BibitemOpen
  \bibfield  {author} {\bibinfo {author} {\bibfnamefont {D.~E.}\ \bibnamefont
  {Chang}}, \bibinfo {author} {\bibfnamefont {J.~S.}\ \bibnamefont {Douglas}},
  \bibinfo {author} {\bibfnamefont {A.}~\bibnamefont {Gonz{\'{a}}lez-Tudela}},
  \bibinfo {author} {\bibfnamefont {C.-L.}\ \bibnamefont {Hung}},\ and\
  \bibinfo {author} {\bibfnamefont {H.~J.}\ \bibnamefont {Kimble}},\ }\bibfield
   {title} {\bibinfo {title} {{Colloquium: Quantum matter built from nanoscopic
  lattices of atoms and photons}},\ }\href
  {https://doi.org/10.1103/RevModPhys.90.031002} {\bibfield  {journal}
  {\bibinfo  {journal} {Reviews of Modern Physics}\ }\textbf {\bibinfo {volume}
  {90}},\ \bibinfo {pages} {031002} (\bibinfo {year} {2018})}\BibitemShut
  {NoStop}%
\bibitem [{\citenamefont {Yu}\ \emph {et~al.}(2019)\citenamefont {Yu},
  \citenamefont {Muniz}, \citenamefont {Hung},\ and\ \citenamefont
  {Kimble}}]{Yu2019}%
  \BibitemOpen
  \bibfield  {author} {\bibinfo {author} {\bibfnamefont {S.-P.}\ \bibnamefont
  {Yu}}, \bibinfo {author} {\bibfnamefont {J.~A.}\ \bibnamefont {Muniz}},
  \bibinfo {author} {\bibfnamefont {C.-L.}\ \bibnamefont {Hung}},\ and\
  \bibinfo {author} {\bibfnamefont {H.~J.}\ \bibnamefont {Kimble}},\ }\bibfield
   {title} {\bibinfo {title} {{Two-dimensional photonic crystals for
  engineering atom–light interactions}},\ }\href
  {https://doi.org/10.1073/pnas.1822110116} {\bibfield  {journal} {\bibinfo
  {journal} {Proceedings of the National Academy of Sciences}\ }\textbf
  {\bibinfo {volume} {116}},\ \bibinfo {pages} {12743} (\bibinfo {year}
  {2019})}\BibitemShut {NoStop}%
\bibitem [{\citenamefont {Shi}\ \emph {et~al.}(2016)\citenamefont {Shi},
  \citenamefont {Wu}, \citenamefont {Gonz\'alez-Tudela},\ and\ \citenamefont
  {Cirac}}]{PhysRevX.6.021027}%
  \BibitemOpen
  \bibfield  {author} {\bibinfo {author} {\bibfnamefont {T.}~\bibnamefont
  {Shi}}, \bibinfo {author} {\bibfnamefont {Y.-H.}\ \bibnamefont {Wu}},
  \bibinfo {author} {\bibfnamefont {A.}~\bibnamefont {Gonz\'alez-Tudela}},\
  and\ \bibinfo {author} {\bibfnamefont {J.~I.}\ \bibnamefont {Cirac}},\
  }\bibfield  {title} {\bibinfo {title} {Bound states in boson impurity
  models},\ }\href {https://doi.org/10.1103/PhysRevX.6.021027} {\bibfield
  {journal} {\bibinfo  {journal} {Phys. Rev. X}\ }\textbf {\bibinfo {volume}
  {6}},\ \bibinfo {pages} {021027} (\bibinfo {year} {2016})}\BibitemShut
  {NoStop}%
\bibitem [{\citenamefont {Calaj\'o}\ \emph {et~al.}(2016)\citenamefont
  {Calaj\'o}, \citenamefont {Ciccarello}, \citenamefont {Chang},\ and\
  \citenamefont {Rabl}}]{PhysRevA.93.033833}%
  \BibitemOpen
  \bibfield  {author} {\bibinfo {author} {\bibfnamefont {G.}~\bibnamefont
  {Calaj\'o}}, \bibinfo {author} {\bibfnamefont {F.}~\bibnamefont
  {Ciccarello}}, \bibinfo {author} {\bibfnamefont {D.}~\bibnamefont {Chang}},\
  and\ \bibinfo {author} {\bibfnamefont {P.}~\bibnamefont {Rabl}},\ }\bibfield
  {title} {\bibinfo {title} {Atom-field dressed states in slow-light waveguide
  qed},\ }\href {https://doi.org/10.1103/PhysRevA.93.033833} {\bibfield
  {journal} {\bibinfo  {journal} {Phys. Rev. A}\ }\textbf {\bibinfo {volume}
  {93}},\ \bibinfo {pages} {033833} (\bibinfo {year} {2016})}\BibitemShut
  {NoStop}%
\bibitem [{\citenamefont {de~Vega}\ \emph {et~al.}(2008)\citenamefont
  {de~Vega}, \citenamefont {Porras},\ and\ \citenamefont
  {Ignacio~Cirac}}]{de_Vega_2008}%
  \BibitemOpen
  \bibfield  {author} {\bibinfo {author} {\bibfnamefont {I.}~\bibnamefont
  {de~Vega}}, \bibinfo {author} {\bibfnamefont {D.}~\bibnamefont {Porras}},\
  and\ \bibinfo {author} {\bibfnamefont {J.}~\bibnamefont {Ignacio~Cirac}},\
  }\bibfield  {title} {\bibinfo {title} {Matter-wave emission in optical
  lattices: Single particle and collective effects},\ }\href
  {http://dx.doi.org/10.1103/PhysRevLett.101.260404} {\bibfield  {journal}
  {\bibinfo  {journal} {Physical Review Letters}\ }\textbf {\bibinfo {volume}
  {101}},\ \bibinfo {pages} {260404} (\bibinfo {year} {2008})}\BibitemShut
  {NoStop}%
\bibitem [{\citenamefont {B{\'{e}}guin}\ \emph {et~al.}(2020)\citenamefont
  {B{\'{e}}guin}, \citenamefont {Laurat}, \citenamefont {Luan}, \citenamefont
  {Burgers}, \citenamefont {Qin},\ and\ \citenamefont {Kimble}}]{Beguin2020}%
  \BibitemOpen
  \bibfield  {author} {\bibinfo {author} {\bibfnamefont {J.-B.}\ \bibnamefont
  {B{\'{e}}guin}}, \bibinfo {author} {\bibfnamefont {J.}~\bibnamefont
  {Laurat}}, \bibinfo {author} {\bibfnamefont {X.}~\bibnamefont {Luan}},
  \bibinfo {author} {\bibfnamefont {A.~P.}\ \bibnamefont {Burgers}}, \bibinfo
  {author} {\bibfnamefont {Z.}~\bibnamefont {Qin}},\ and\ \bibinfo {author}
  {\bibfnamefont {H.~J.}\ \bibnamefont {Kimble}},\ }\bibfield  {title}
  {\bibinfo {title} {{Reduced volume and reflection for bright optical tweezers
  with radial Laguerre–Gauss beams}},\ }\href
  {https://doi.org/10.1073/pnas.2014017117} {\bibfield  {journal} {\bibinfo
  {journal} {Proceedings of the National Academy of Sciences}\ }\textbf
  {\bibinfo {volume} {117}},\ \bibinfo {pages} {26109} (\bibinfo {year}
  {2020})}\BibitemShut {NoStop}%
\bibitem [{\citenamefont {Masson}\ and\ \citenamefont
  {Asenjo-Garcia}(2020)}]{masson2019atomicwaveguide}%
  \BibitemOpen
  \bibfield  {author} {\bibinfo {author} {\bibfnamefont {S.~J.}\ \bibnamefont
  {Masson}}\ and\ \bibinfo {author} {\bibfnamefont {A.}~\bibnamefont
  {Asenjo-Garcia}},\ }\bibfield  {title} {\bibinfo {title} {Atomic-waveguide
  quantum electrodynamics},\ }\href
  {https://doi.org/10.1103/PhysRevResearch.2.043213} {\bibfield  {journal}
  {\bibinfo  {journal} {Phys. Rev. Research}\ }\textbf {\bibinfo {volume}
  {2}},\ \bibinfo {pages} {043213} (\bibinfo {year} {2020})}\BibitemShut
  {NoStop}%
\bibitem [{\citenamefont {Patti}\ \emph {et~al.}(2020)\citenamefont {Patti},
  \citenamefont {Wild}, \citenamefont {Shahmoon}, \citenamefont {Lukin},\ and\
  \citenamefont {Yelin}}]{patti2020controlling}%
  \BibitemOpen
  \bibfield  {author} {\bibinfo {author} {\bibfnamefont {T.~L.}\ \bibnamefont
  {Patti}}, \bibinfo {author} {\bibfnamefont {D.~S.}\ \bibnamefont {Wild}},
  \bibinfo {author} {\bibfnamefont {E.}~\bibnamefont {Shahmoon}}, \bibinfo
  {author} {\bibfnamefont {M.~D.}\ \bibnamefont {Lukin}},\ and\ \bibinfo
  {author} {\bibfnamefont {S.~F.}\ \bibnamefont {Yelin}},\ }\href@noop {}
  {\bibinfo {title} {Controlling interactions between quantum emitters using
  atom arrays}} (\bibinfo {year} {2020}),\ \Eprint
  {https://arxiv.org/abs/2005.03495} {arXiv:2005.03495 [quant-ph]} \BibitemShut
  {NoStop}%
\bibitem [{\citenamefont {Bettles}\ \emph {et~al.}(2017)\citenamefont
  {Bettles}, \citenamefont {Min{\'{a}}ř}, \citenamefont {Adams}, \citenamefont
  {Lesanovsky},\ and\ \citenamefont {Olmos}}]{Bettles2017}%
  \BibitemOpen
  \bibfield  {author} {\bibinfo {author} {\bibfnamefont {R.~J.}\ \bibnamefont
  {Bettles}}, \bibinfo {author} {\bibfnamefont {J.}~\bibnamefont
  {Min{\'{a}}ř}}, \bibinfo {author} {\bibfnamefont {C.~S.}\ \bibnamefont
  {Adams}}, \bibinfo {author} {\bibfnamefont {I.}~\bibnamefont {Lesanovsky}},\
  and\ \bibinfo {author} {\bibfnamefont {B.}~\bibnamefont {Olmos}},\ }\bibfield
   {title} {\bibinfo {title} {{Topological properties of a dense atomic lattice
  gas}},\ }\href {https://doi.org/10.1103/PhysRevA.96.041603} {\bibfield
  {journal} {\bibinfo  {journal} {Physical Review A}\ }\textbf {\bibinfo
  {volume} {96}},\ \bibinfo {pages} {041603} (\bibinfo {year}
  {2017})}\BibitemShut {NoStop}%
\bibitem [{\citenamefont {Perczel}\ \emph
  {et~al.}(2017{\natexlab{a}})\citenamefont {Perczel}, \citenamefont
  {Borregaard}, \citenamefont {Chang}, \citenamefont {Pichler}, \citenamefont
  {Yelin}, \citenamefont {Zoller},\ and\ \citenamefont {Lukin}}]{Perczel2017}%
  \BibitemOpen
  \bibfield  {author} {\bibinfo {author} {\bibfnamefont {J.}~\bibnamefont
  {Perczel}}, \bibinfo {author} {\bibfnamefont {J.}~\bibnamefont {Borregaard}},
  \bibinfo {author} {\bibfnamefont {D.~E.}\ \bibnamefont {Chang}}, \bibinfo
  {author} {\bibfnamefont {H.}~\bibnamefont {Pichler}}, \bibinfo {author}
  {\bibfnamefont {S.~F.}\ \bibnamefont {Yelin}}, \bibinfo {author}
  {\bibfnamefont {P.}~\bibnamefont {Zoller}},\ and\ \bibinfo {author}
  {\bibfnamefont {M.~D.}\ \bibnamefont {Lukin}},\ }\bibfield  {title} {\bibinfo
  {title} {{Topological Quantum Optics in Two-Dimensional Atomic Arrays}},\
  }\href {https://doi.org/10.1103/PhysRevLett.119.023603} {\bibfield  {journal}
  {\bibinfo  {journal} {Physical Review Letters}\ }\textbf {\bibinfo {volume}
  {119}},\ \bibinfo {pages} {023603} (\bibinfo {year}
  {2017}{\natexlab{a}})}\BibitemShut {NoStop}%
\bibitem [{\citenamefont {Perczel}\ \emph
  {et~al.}(2017{\natexlab{b}})\citenamefont {Perczel}, \citenamefont
  {Borregaard}, \citenamefont {Chang}, \citenamefont {Pichler}, \citenamefont
  {Yelin}, \citenamefont {Zoller},\ and\ \citenamefont {Lukin}}]{Perczel2017a}%
  \BibitemOpen
  \bibfield  {author} {\bibinfo {author} {\bibfnamefont {J.}~\bibnamefont
  {Perczel}}, \bibinfo {author} {\bibfnamefont {J.}~\bibnamefont {Borregaard}},
  \bibinfo {author} {\bibfnamefont {D.~E.}\ \bibnamefont {Chang}}, \bibinfo
  {author} {\bibfnamefont {H.}~\bibnamefont {Pichler}}, \bibinfo {author}
  {\bibfnamefont {S.~F.}\ \bibnamefont {Yelin}}, \bibinfo {author}
  {\bibfnamefont {P.}~\bibnamefont {Zoller}},\ and\ \bibinfo {author}
  {\bibfnamefont {M.~D.}\ \bibnamefont {Lukin}},\ }\bibfield  {title} {\bibinfo
  {title} {{Photonic band structure of two-dimensional atomic lattices}},\
  }\href {https://doi.org/10.1103/PhysRevA.96.063801} {\bibfield  {journal}
  {\bibinfo  {journal} {Physical Review A}\ }\textbf {\bibinfo {volume} {96}},\
  \bibinfo {pages} {063801} (\bibinfo {year} {2017}{\natexlab{b}})}\BibitemShut
  {NoStop}%
\bibitem [{\citenamefont {Jenkins}\ and\ \citenamefont
  {Ruostekoski}(2013)}]{Jenkins2013}%
  \BibitemOpen
  \bibfield  {author} {\bibinfo {author} {\bibfnamefont {S.~D.}\ \bibnamefont
  {Jenkins}}\ and\ \bibinfo {author} {\bibfnamefont {J.}~\bibnamefont
  {Ruostekoski}},\ }\bibfield  {title} {\bibinfo {title} {{Metamaterial
  Transparency Induced by Cooperative Electromagnetic Interactions}},\ }\href
  {https://doi.org/10.1103/PhysRevLett.111.147401} {\bibfield  {journal}
  {\bibinfo  {journal} {Physical Review Letters}\ }\textbf {\bibinfo {volume}
  {111}},\ \bibinfo {pages} {147401} (\bibinfo {year} {2013})}\BibitemShut
  {NoStop}%
\bibitem [{\citenamefont {Bettles}\ \emph {et~al.}(2016)\citenamefont
  {Bettles}, \citenamefont {Gardiner},\ and\ \citenamefont
  {Adams}}]{Bettles2016}%
  \BibitemOpen
  \bibfield  {author} {\bibinfo {author} {\bibfnamefont {R.~J.}\ \bibnamefont
  {Bettles}}, \bibinfo {author} {\bibfnamefont {S.~A.}\ \bibnamefont
  {Gardiner}},\ and\ \bibinfo {author} {\bibfnamefont {C.~S.}\ \bibnamefont
  {Adams}},\ }\bibfield  {title} {\bibinfo {title} {{Enhanced Optical Cross
  Section via Collective Coupling of Atomic Dipoles in a 2D Array}},\ }\href
  {https://doi.org/10.1103/PhysRevLett.116.103602} {\bibfield  {journal}
  {\bibinfo  {journal} {Physical Review Letters}\ }\textbf {\bibinfo {volume}
  {116}},\ \bibinfo {pages} {103602} (\bibinfo {year} {2016})}\BibitemShut
  {NoStop}%
\bibitem [{\citenamefont {Shahmoon}\ \emph {et~al.}(2017)\citenamefont
  {Shahmoon}, \citenamefont {Wild}, \citenamefont {Lukin},\ and\ \citenamefont
  {Yelin}}]{Shahmoon2017}%
  \BibitemOpen
  \bibfield  {author} {\bibinfo {author} {\bibfnamefont {E.}~\bibnamefont
  {Shahmoon}}, \bibinfo {author} {\bibfnamefont {D.~S.}\ \bibnamefont {Wild}},
  \bibinfo {author} {\bibfnamefont {M.~D.}\ \bibnamefont {Lukin}},\ and\
  \bibinfo {author} {\bibfnamefont {S.~F.}\ \bibnamefont {Yelin}},\ }\bibfield
  {title} {\bibinfo {title} {{Cooperative Resonances in Light Scattering from
  Two-Dimensional Atomic Arrays}},\ }\href
  {https://doi.org/10.1103/PhysRevLett.118.113601} {\bibfield  {journal}
  {\bibinfo  {journal} {Physical Review Letters}\ }\textbf {\bibinfo {volume}
  {118}},\ \bibinfo {pages} {113601} (\bibinfo {year} {2017})}\BibitemShut
  {NoStop}%
\bibitem [{\citenamefont {Asenjo-Garcia}\ \emph
  {et~al.}(2017{\natexlab{a}})\citenamefont {Asenjo-Garcia}, \citenamefont
  {Moreno-Cardoner}, \citenamefont {Albrecht}, \citenamefont {Kimble},\ and\
  \citenamefont {Chang}}]{Asenjo-Garcia2017}%
  \BibitemOpen
  \bibfield  {author} {\bibinfo {author} {\bibfnamefont {A.}~\bibnamefont
  {Asenjo-Garcia}}, \bibinfo {author} {\bibfnamefont {M.}~\bibnamefont
  {Moreno-Cardoner}}, \bibinfo {author} {\bibfnamefont {A.}~\bibnamefont
  {Albrecht}}, \bibinfo {author} {\bibfnamefont {H.~J.}\ \bibnamefont
  {Kimble}},\ and\ \bibinfo {author} {\bibfnamefont {D.~E.}\ \bibnamefont
  {Chang}},\ }\bibfield  {title} {\bibinfo {title} {{Exponential Improvement in
  Photon Storage Fidelities Using Subradiance and “Selective Radiance” in
  Atomic Arrays}},\ }\href {https://doi.org/10.1103/PhysRevX.7.031024}
  {\bibfield  {journal} {\bibinfo  {journal} {Physical Review X}\ }\textbf
  {\bibinfo {volume} {7}},\ \bibinfo {pages} {031024} (\bibinfo {year}
  {2017}{\natexlab{a}})}\BibitemShut {NoStop}%
\bibitem [{\citenamefont {Shahmoon}\ \emph {et~al.}(2020)\citenamefont
  {Shahmoon}, \citenamefont {Lukin},\ and\ \citenamefont
  {Yelin}}]{Shahmoon2020}%
  \BibitemOpen
  \bibfield  {author} {\bibinfo {author} {\bibfnamefont {E.}~\bibnamefont
  {Shahmoon}}, \bibinfo {author} {\bibfnamefont {M.~D.}\ \bibnamefont
  {Lukin}},\ and\ \bibinfo {author} {\bibfnamefont {S.~F.}\ \bibnamefont
  {Yelin}},\ }\bibfield  {title} {\bibinfo {title} {{Quantum optomechanics of a
  two-dimensional atomic array}},\ }\href
  {https://doi.org/10.1103/PhysRevA.101.063833} {\bibfield  {journal} {\bibinfo
   {journal} {Physical Review A}\ }\textbf {\bibinfo {volume} {101}},\ \bibinfo
  {pages} {063833} (\bibinfo {year} {2020})}\BibitemShut {NoStop}%
\bibitem [{\citenamefont {Rui}\ \emph {et~al.}(2020)\citenamefont {Rui},
  \citenamefont {Wei}, \citenamefont {Rubio-Abadal}, \citenamefont {Hollerith},
  \citenamefont {Zeiher}, \citenamefont {Stamper-Kurn}, \citenamefont {Gross},\
  and\ \citenamefont {Bloch}}]{Rui2020}%
  \BibitemOpen
  \bibfield  {author} {\bibinfo {author} {\bibfnamefont {J.}~\bibnamefont
  {Rui}}, \bibinfo {author} {\bibfnamefont {D.}~\bibnamefont {Wei}}, \bibinfo
  {author} {\bibfnamefont {A.}~\bibnamefont {Rubio-Abadal}}, \bibinfo {author}
  {\bibfnamefont {S.}~\bibnamefont {Hollerith}}, \bibinfo {author}
  {\bibfnamefont {J.}~\bibnamefont {Zeiher}}, \bibinfo {author} {\bibfnamefont
  {D.~M.}\ \bibnamefont {Stamper-Kurn}}, \bibinfo {author} {\bibfnamefont
  {C.}~\bibnamefont {Gross}},\ and\ \bibinfo {author} {\bibfnamefont
  {I.}~\bibnamefont {Bloch}},\ }\bibfield  {title} {\bibinfo {title} {{A
  subradiant optical mirror formed by a single structured atomic layer}},\
  }\href {https://doi.org/10.1038/s41586-020-2463-x} {\bibfield  {journal}
  {\bibinfo  {journal} {Nature}\ }\textbf {\bibinfo {volume} {583}},\ \bibinfo
  {pages} {369} (\bibinfo {year} {2020})}\BibitemShut {NoStop}%
\bibitem [{\citenamefont {Melzer}\ and\ \citenamefont
  {McLeod}(2020)}]{3Dnanophotonic}%
  \BibitemOpen
  \bibfield  {author} {\bibinfo {author} {\bibfnamefont {J.~E.}\ \bibnamefont
  {Melzer}}\ and\ \bibinfo {author} {\bibfnamefont {E.}~\bibnamefont
  {McLeod}},\ }\bibfield  {title} {\bibinfo {title} {3d nanophotonic device
  fabrication using discrete components},\ }\href
  {https://doi.org/https://doi.org/10.1515/nanoph-2020-0161} {\bibfield
  {journal} {\bibinfo  {journal} {Nanophotonics}\ }\textbf {\bibinfo {volume}
  {9}},\ \bibinfo {pages} {1373 } (\bibinfo {year} {01 Jun. 2020})}\BibitemShut
  {NoStop}%
\bibitem [{\citenamefont {Bradac}\ \emph {et~al.}(2019)\citenamefont {Bradac},
  \citenamefont {Gao}, \citenamefont {Forneris}, \citenamefont {Trusheim},\
  and\ \citenamefont {Aharonovich}}]{Bradac2019}%
  \BibitemOpen
  \bibfield  {author} {\bibinfo {author} {\bibfnamefont {C.}~\bibnamefont
  {Bradac}}, \bibinfo {author} {\bibfnamefont {W.}~\bibnamefont {Gao}},
  \bibinfo {author} {\bibfnamefont {J.}~\bibnamefont {Forneris}}, \bibinfo
  {author} {\bibfnamefont {M.~E.}\ \bibnamefont {Trusheim}},\ and\ \bibinfo
  {author} {\bibfnamefont {I.}~\bibnamefont {Aharonovich}},\ }\bibfield
  {title} {\bibinfo {title} {{Quantum nanophotonics with group IV defects in
  diamond}},\ }\href {https://doi.org/10.1038/s41467-019-13332-w} {\bibfield
  {journal} {\bibinfo  {journal} {Nature Communications}\ }\textbf {\bibinfo
  {volume} {10}},\ \bibinfo {pages} {5625} (\bibinfo {year}
  {2019})}\BibitemShut {NoStop}%
\bibitem [{\citenamefont {Antezza}\ and\ \citenamefont
  {Castin}(2009{\natexlab{a}})}]{fluctuations}%
  \BibitemOpen
  \bibfield  {author} {\bibinfo {author} {\bibfnamefont {M.}~\bibnamefont
  {Antezza}}\ and\ \bibinfo {author} {\bibfnamefont {Y.}~\bibnamefont
  {Castin}},\ }\bibfield  {title} {\bibinfo {title} {Spectrum of light in a
  quantum fluctuating periodic structure},\ }\href
  {https://doi.org/10.1103/physrevlett.103.123903} {\bibfield  {journal}
  {\bibinfo  {journal} {Physical Review Letters}\ }\textbf {\bibinfo {volume}
  {103}},\ \bibinfo {pages} {123903} (\bibinfo {year}
  {2009}{\natexlab{a}})}\BibitemShut {NoStop}%
\bibitem [{\citenamefont {Klugkist}\ \emph {et~al.}(2006)\citenamefont
  {Klugkist}, \citenamefont {Mostovoy},\ and\ \citenamefont
  {Knoester}}]{Klugkist2006}%
  \BibitemOpen
  \bibfield  {author} {\bibinfo {author} {\bibfnamefont {J.}~\bibnamefont
  {Klugkist}}, \bibinfo {author} {\bibfnamefont {M.}~\bibnamefont {Mostovoy}},\
  and\ \bibinfo {author} {\bibfnamefont {J.}~\bibnamefont {Knoester}},\
  }\bibfield  {title} {\bibinfo {title} {Mode softening, ferroelectric
  transition, and tunable photonic band structures in a point-dipole crystal},\
  }\href {https://doi.org/10.1103/PhysRevLett.96.163903} {\bibfield  {journal}
  {\bibinfo  {journal} {Physical review letters}\ }\textbf {\bibinfo {volume}
  {96}},\ \bibinfo {pages} {163903} (\bibinfo {year} {2006})}\BibitemShut
  {NoStop}%
\bibitem [{\citenamefont {Deutsch}\ \emph {et~al.}(1995)\citenamefont
  {Deutsch}, \citenamefont {Spreeuw}, \citenamefont {Rolston},\ and\
  \citenamefont {Phillips}}]{PhysRevA.52.1394}%
  \BibitemOpen
  \bibfield  {author} {\bibinfo {author} {\bibfnamefont {I.~H.}\ \bibnamefont
  {Deutsch}}, \bibinfo {author} {\bibfnamefont {R.~J.~C.}\ \bibnamefont
  {Spreeuw}}, \bibinfo {author} {\bibfnamefont {S.~L.}\ \bibnamefont
  {Rolston}},\ and\ \bibinfo {author} {\bibfnamefont {W.~D.}\ \bibnamefont
  {Phillips}},\ }\bibfield  {title} {\bibinfo {title} {Photonic band gaps in
  optical lattices},\ }\href {https://doi.org/10.1103/PhysRevA.52.1394}
  {\bibfield  {journal} {\bibinfo  {journal} {Phys. Rev. A}\ }\textbf {\bibinfo
  {volume} {52}},\ \bibinfo {pages} {1394} (\bibinfo {year}
  {1995})}\BibitemShut {NoStop}%
\bibitem [{\citenamefont {van Coevorden}\ \emph {et~al.}(1996)\citenamefont
  {van Coevorden}, \citenamefont {Sprik}, \citenamefont {Tip},\ and\
  \citenamefont {Lagendijk}}]{PhysRevLett.77.2412}%
  \BibitemOpen
  \bibfield  {author} {\bibinfo {author} {\bibfnamefont {D.~V.}\ \bibnamefont
  {van Coevorden}}, \bibinfo {author} {\bibfnamefont {R.}~\bibnamefont
  {Sprik}}, \bibinfo {author} {\bibfnamefont {A.}~\bibnamefont {Tip}},\ and\
  \bibinfo {author} {\bibfnamefont {A.}~\bibnamefont {Lagendijk}},\ }\bibfield
  {title} {\bibinfo {title} {Photonic band structure of atomic lattices},\
  }\href {https://doi.org/10.1103/PhysRevLett.77.2412} {\bibfield  {journal}
  {\bibinfo  {journal} {Phys. Rev. Lett.}\ }\textbf {\bibinfo {volume} {77}},\
  \bibinfo {pages} {2412} (\bibinfo {year} {1996})}\BibitemShut {NoStop}%
\bibitem [{\citenamefont {Yu}(2011)}]{PhysRevA.84.043833}%
  \BibitemOpen
  \bibfield  {author} {\bibinfo {author} {\bibfnamefont {D.}~\bibnamefont
  {Yu}},\ }\bibfield  {title} {\bibinfo {title} {Photonic band structure of the
  three-dimensional ${}^{88}\mathbf{Sr}$ atomic lattice},\ }\href
  {https://doi.org/10.1103/PhysRevA.84.043833} {\bibfield  {journal} {\bibinfo
  {journal} {Phys. Rev. A}\ }\textbf {\bibinfo {volume} {84}},\ \bibinfo
  {pages} {043833} (\bibinfo {year} {2011})}\BibitemShut {NoStop}%
\bibitem [{\citenamefont {Antezza}\ and\ \citenamefont
  {Castin}(2009{\natexlab{b}})}]{antezza}%
  \BibitemOpen
  \bibfield  {author} {\bibinfo {author} {\bibfnamefont {M.}~\bibnamefont
  {Antezza}}\ and\ \bibinfo {author} {\bibfnamefont {Y.}~\bibnamefont
  {Castin}},\ }\bibfield  {title} {\bibinfo {title} {Fano-hopfield model and
  photonic band gaps for an arbitrary atomic lattice},\ }\href@noop {}
  {\bibfield  {journal} {\bibinfo  {journal} {Phys. Rev. A}\ }\textbf {\bibinfo
  {volume} {80}},\ \bibinfo {pages} {013816} (\bibinfo {year}
  {2009}{\natexlab{b}})}\BibitemShut {NoStop}%
\bibitem [{\citenamefont {Greiner}\ \emph {et~al.}(2002)\citenamefont
  {Greiner}, \citenamefont {Mandel}, \citenamefont {Esslinger}, \citenamefont
  {H{\"{a}}nsch},\ and\ \citenamefont {Bloch}}]{Greiner2002}%
  \BibitemOpen
  \bibfield  {author} {\bibinfo {author} {\bibfnamefont {M.}~\bibnamefont
  {Greiner}}, \bibinfo {author} {\bibfnamefont {O.}~\bibnamefont {Mandel}},
  \bibinfo {author} {\bibfnamefont {T.}~\bibnamefont {Esslinger}}, \bibinfo
  {author} {\bibfnamefont {T.~W.}\ \bibnamefont {H{\"{a}}nsch}},\ and\ \bibinfo
  {author} {\bibfnamefont {I.}~\bibnamefont {Bloch}},\ }\bibfield  {title}
  {\bibinfo {title} {{Quantum phase transition from a superfluid to a Mott
  insulator in a gas of ultracold atoms}},\ }\href
  {https://doi.org/10.1038/415039a} {\bibfield  {journal} {\bibinfo  {journal}
  {Nature}\ }\textbf {\bibinfo {volume} {415}},\ \bibinfo {pages} {39}
  (\bibinfo {year} {2002})}\BibitemShut {NoStop}%
\bibitem [{\citenamefont {K\"ohl}\ \emph {et~al.}(2005)\citenamefont {K\"ohl},
  \citenamefont {Moritz}, \citenamefont {St\"oferle}, \citenamefont
  {G\"unter},\ and\ \citenamefont {Esslinger}}]{PhysRevLett.94.080403}%
  \BibitemOpen
  \bibfield  {author} {\bibinfo {author} {\bibfnamefont {M.}~\bibnamefont
  {K\"ohl}}, \bibinfo {author} {\bibfnamefont {H.}~\bibnamefont {Moritz}},
  \bibinfo {author} {\bibfnamefont {T.}~\bibnamefont {St\"oferle}}, \bibinfo
  {author} {\bibfnamefont {K.}~\bibnamefont {G\"unter}},\ and\ \bibinfo
  {author} {\bibfnamefont {T.}~\bibnamefont {Esslinger}},\ }\bibfield  {title}
  {\bibinfo {title} {Fermionic atoms in a three dimensional optical lattice:
  Observing fermi surfaces, dynamics, and interactions},\ }\href
  {https://doi.org/10.1103/PhysRevLett.94.080403} {\bibfield  {journal}
  {\bibinfo  {journal} {Phys. Rev. Lett.}\ }\textbf {\bibinfo {volume} {94}},\
  \bibinfo {pages} {080403} (\bibinfo {year} {2005})}\BibitemShut {NoStop}%
\bibitem [{\citenamefont {Lehmberg}(1970{\natexlab{a}})}]{Lehmberg1970}%
  \BibitemOpen
  \bibfield  {author} {\bibinfo {author} {\bibfnamefont {R.~H.}\ \bibnamefont
  {Lehmberg}},\ }\bibfield  {title} {\bibinfo {title} {{Radiation from an
  N-Atom System. I. General Formalism}},\ }\href
  {https://doi.org/10.1103/PhysRevA.2.883} {\bibfield  {journal} {\bibinfo
  {journal} {Physical Review A}\ }\textbf {\bibinfo {volume} {2}},\ \bibinfo
  {pages} {883} (\bibinfo {year} {1970}{\natexlab{a}})}\BibitemShut {NoStop}%
\bibitem [{\citenamefont {Lehmberg}(1970{\natexlab{b}})}]{Lehmberg1970a}%
  \BibitemOpen
  \bibfield  {author} {\bibinfo {author} {\bibfnamefont {R.~H.}\ \bibnamefont
  {Lehmberg}},\ }\bibfield  {title} {\bibinfo {title} {{Radiation from an
  N-Atom System. II. Spontaneous Emission from a Pair of Atoms}},\ }\href
  {https://doi.org/10.1103/PhysRevA.2.889} {\bibfield  {journal} {\bibinfo
  {journal} {Physical Review A}\ }\textbf {\bibinfo {volume} {2}},\ \bibinfo
  {pages} {889} (\bibinfo {year} {1970}{\natexlab{b}})}\BibitemShut {NoStop}%
\bibitem [{\citenamefont {Dung}\ \emph {et~al.}(1998)\citenamefont {Dung},
  \citenamefont {Kn\"oll},\ and\ \citenamefont {Welsch}}]{PhysRevA.57.3931}%
  \BibitemOpen
  \bibfield  {author} {\bibinfo {author} {\bibfnamefont {H.~T.}\ \bibnamefont
  {Dung}}, \bibinfo {author} {\bibfnamefont {L.}~\bibnamefont {Kn\"oll}},\ and\
  \bibinfo {author} {\bibfnamefont {D.-G.}\ \bibnamefont {Welsch}},\ }\bibfield
   {title} {\bibinfo {title} {Three-dimensional quantization of the
  electromagnetic field in dispersive and absorbing inhomogeneous
  dielectrics},\ }\href {https://doi.org/10.1103/PhysRevA.57.3931} {\bibfield
  {journal} {\bibinfo  {journal} {Phys. Rev. A}\ }\textbf {\bibinfo {volume}
  {57}},\ \bibinfo {pages} {3931} (\bibinfo {year} {1998})}\BibitemShut
  {NoStop}%
\bibitem [{\citenamefont {Perczel}\ \emph
  {et~al.}(2017{\natexlab{c}})\citenamefont {Perczel}, \citenamefont
  {Borregaard}, \citenamefont {Chang}, \citenamefont {Pichler}, \citenamefont
  {Yelin}, \citenamefont {Zoller},\ and\ \citenamefont
  {Lukin}}]{PhysRevA.96.063801}%
  \BibitemOpen
  \bibfield  {author} {\bibinfo {author} {\bibfnamefont {J.}~\bibnamefont
  {Perczel}}, \bibinfo {author} {\bibfnamefont {J.}~\bibnamefont {Borregaard}},
  \bibinfo {author} {\bibfnamefont {D.~E.}\ \bibnamefont {Chang}}, \bibinfo
  {author} {\bibfnamefont {H.}~\bibnamefont {Pichler}}, \bibinfo {author}
  {\bibfnamefont {S.~F.}\ \bibnamefont {Yelin}}, \bibinfo {author}
  {\bibfnamefont {P.}~\bibnamefont {Zoller}},\ and\ \bibinfo {author}
  {\bibfnamefont {M.~D.}\ \bibnamefont {Lukin}},\ }\bibfield  {title} {\bibinfo
  {title} {Photonic band structure of two-dimensional atomic lattices},\ }\href
  {https://doi.org/10.1103/PhysRevA.96.063801} {\bibfield  {journal} {\bibinfo
  {journal} {Phys. Rev. A}\ }\textbf {\bibinfo {volume} {96}},\ \bibinfo
  {pages} {063801} (\bibinfo {year} {2017}{\natexlab{c}})}\BibitemShut
  {NoStop}%
\bibitem [{Note1()}]{Note1}%
  \BibitemOpen
  \bibinfo {note} {$\phi ,k_x$ and $k_y$ are determined by the condition
  $\protect \bm {\epsilon } \perp \protect \mathbf {k}$ and the detuning of the
  laser from $\omega _0$.}\BibitemShut {Stop}%
\bibitem [{Note2()}]{Note2}%
  \BibitemOpen
  \bibinfo {note} {The form of \protect \cref {nonbravaisinteraction} can be
  understood intuitively: an atom in the sublattice $A$ is in the excited state
  $\mathinner {|{x}\rangle }$ with dipole momentum $\protect \mathbf {d}=d_0
  (1,0,0)$ may radiate light with momentum $\protect \mathbf {k}_1=(k_x,0,\pi
  /(2a))$ and $ \protect \mathbf {k}_2=(k_x,0,-\pi /(2a))$. The associated
  electric field is $\protect \mathbf {E}(\protect \mathbf {r})\propto \left
  (\pi /(2a)\protect \mathbf {e}_x \protect \qopname \relax o{cos}(\pi
  /(2a)z)-ik_x \protect \mathbf {e}_z \protect \qopname \relax o{sin}(\pi
  /(2a)z)\right )\protect \qopname \relax o{exp}(ik_xx)$. (The asymmetric
  superposition of $e^{i\protect \mathbf {k}_1\protect \mathbf {r}}$ and
  $e^{i\protect \mathbf {k}_2\protect \mathbf {r}}$ is zero at the position of
  the emitting atom and thus decoupled.) Hence, the $x$-polarized atoms in
  sublattice $A$ couple to the $z$-polarized atoms in sublattice $B$ (but not
  to the $z$-polarization of sublattice $A$) and vice versa. The same argument
  can be repeated for $y$-polarization.}\BibitemShut {Stop}%
\bibitem [{\citenamefont {Antezza}\ and\ \citenamefont
  {Castin}(2013)}]{antezza2013}%
  \BibitemOpen
  \bibfield  {author} {\bibinfo {author} {\bibfnamefont {M.}~\bibnamefont
  {Antezza}}\ and\ \bibinfo {author} {\bibfnamefont {Y.}~\bibnamefont
  {Castin}},\ }\bibfield  {title} {\bibinfo {title} {Photonic band gap in an
  imperfect atomic diamond lattice: Penetration depth and effects of finite
  size and vacancies},\ }\href {https://doi.org/10.1103/PhysRevA.88.033844}
  {\bibfield  {journal} {\bibinfo  {journal} {Phys. Rev. A}\ }\textbf {\bibinfo
  {volume} {88}},\ \bibinfo {pages} {033844} (\bibinfo {year}
  {2013})}\BibitemShut {NoStop}%
\bibitem [{\citenamefont {Asenjo-Garcia}\ \emph
  {et~al.}(2017{\natexlab{b}})\citenamefont {Asenjo-Garcia}, \citenamefont
  {Moreno-Cardoner}, \citenamefont {Albrecht}, \citenamefont {Kimble},\ and\
  \citenamefont {Chang}}]{PhysRevX.7.031024}%
  \BibitemOpen
  \bibfield  {author} {\bibinfo {author} {\bibfnamefont {A.}~\bibnamefont
  {Asenjo-Garcia}}, \bibinfo {author} {\bibfnamefont {M.}~\bibnamefont
  {Moreno-Cardoner}}, \bibinfo {author} {\bibfnamefont {A.}~\bibnamefont
  {Albrecht}}, \bibinfo {author} {\bibfnamefont {H.~J.}\ \bibnamefont
  {Kimble}},\ and\ \bibinfo {author} {\bibfnamefont {D.~E.}\ \bibnamefont
  {Chang}},\ }\bibfield  {title} {\bibinfo {title} {Exponential improvement in
  photon storage fidelities using subradiance and ``selective radiance'' in
  atomic arrays},\ }\href {https://doi.org/10.1103/PhysRevX.7.031024}
  {\bibfield  {journal} {\bibinfo  {journal} {Phys. Rev. X}\ }\textbf {\bibinfo
  {volume} {7}},\ \bibinfo {pages} {031024} (\bibinfo {year}
  {2017}{\natexlab{b}})}\BibitemShut {NoStop}%
\bibitem [{\citenamefont {Tomita}\ \emph {et~al.}(2017)\citenamefont {Tomita},
  \citenamefont {Nakajima}, \citenamefont {Danshita}, \citenamefont {Takasu},\
  and\ \citenamefont {Takahashi}}]{Tomitae1701513}%
  \BibitemOpen
  \bibfield  {author} {\bibinfo {author} {\bibfnamefont {T.}~\bibnamefont
  {Tomita}}, \bibinfo {author} {\bibfnamefont {S.}~\bibnamefont {Nakajima}},
  \bibinfo {author} {\bibfnamefont {I.}~\bibnamefont {Danshita}}, \bibinfo
  {author} {\bibfnamefont {Y.}~\bibnamefont {Takasu}},\ and\ \bibinfo {author}
  {\bibfnamefont {Y.}~\bibnamefont {Takahashi}},\ }\bibfield  {title} {\bibinfo
  {title} {Observation of the mott insulator to superfluid crossover of a
  driven-dissipative bose-hubbard system},\ }\href
  {https://doi.org/10.1126/sciadv.1701513} {\bibfield  {journal} {\bibinfo
  {journal} {Science Advances}\ }\textbf {\bibinfo {volume} {3}},\ \bibinfo
  {pages} {e1701513} (\bibinfo {year} {2017})}\BibitemShut {NoStop}%
\bibitem [{\citenamefont {Zhao}\ \emph {et~al.}(2009)\citenamefont {Zhao},
  \citenamefont {Chen}, \citenamefont {Bao}, \citenamefont {Strassel},
  \citenamefont {Chuu}, \citenamefont {Jin}, \citenamefont {Schmiedmayer},
  \citenamefont {Yuan}, \citenamefont {Chen},\ and\ \citenamefont
  {Pan}}]{Zhao2009}%
  \BibitemOpen
  \bibfield  {author} {\bibinfo {author} {\bibfnamefont {B.}~\bibnamefont
  {Zhao}}, \bibinfo {author} {\bibfnamefont {Y.-A.}\ \bibnamefont {Chen}},
  \bibinfo {author} {\bibfnamefont {X.-H.}\ \bibnamefont {Bao}}, \bibinfo
  {author} {\bibfnamefont {T.}~\bibnamefont {Strassel}}, \bibinfo {author}
  {\bibfnamefont {C.-S.}\ \bibnamefont {Chuu}}, \bibinfo {author}
  {\bibfnamefont {X.-M.}\ \bibnamefont {Jin}}, \bibinfo {author} {\bibfnamefont
  {J.}~\bibnamefont {Schmiedmayer}}, \bibinfo {author} {\bibfnamefont {Z.-S.}\
  \bibnamefont {Yuan}}, \bibinfo {author} {\bibfnamefont {S.}~\bibnamefont
  {Chen}},\ and\ \bibinfo {author} {\bibfnamefont {J.-W.}\ \bibnamefont
  {Pan}},\ }\bibfield  {title} {\bibinfo {title} {{A millisecond quantum memory
  for scalable quantum networks}},\ }\href {https://doi.org/10.1038/nphys1153}
  {\bibfield  {journal} {\bibinfo  {journal} {Nature Physics}\ }\textbf
  {\bibinfo {volume} {5}},\ \bibinfo {pages} {95} (\bibinfo {year}
  {2009})}\BibitemShut {NoStop}%
\bibitem [{\citenamefont {Jaksch}\ and\ \citenamefont
  {Zoller}(2005)}]{Jaksch2005}%
  \BibitemOpen
  \bibfield  {author} {\bibinfo {author} {\bibfnamefont {D.}~\bibnamefont
  {Jaksch}}\ and\ \bibinfo {author} {\bibfnamefont {P.}~\bibnamefont
  {Zoller}},\ }\bibfield  {title} {\bibinfo {title} {{The cold atom Hubbard
  toolbox}},\ }\href {https://doi.org/10.1016/j.aop.2004.09.010} {\bibfield
  {journal} {\bibinfo  {journal} {Annals of Physics}\ }\textbf {\bibinfo
  {volume} {315}},\ \bibinfo {pages} {52} (\bibinfo {year} {2005})}\BibitemShut
  {NoStop}%
\bibitem [{\citenamefont {Schneider}\ \emph {et~al.}(2008)\citenamefont
  {Schneider}, \citenamefont {Hackermuller}, \citenamefont {Will},
  \citenamefont {Best}, \citenamefont {Bloch}, \citenamefont {Costi},
  \citenamefont {Helmes}, \citenamefont {Rasch},\ and\ \citenamefont
  {Rosch}}]{Schneider2008}%
  \BibitemOpen
  \bibfield  {author} {\bibinfo {author} {\bibfnamefont {U.}~\bibnamefont
  {Schneider}}, \bibinfo {author} {\bibfnamefont {L.}~\bibnamefont
  {Hackermuller}}, \bibinfo {author} {\bibfnamefont {S.}~\bibnamefont {Will}},
  \bibinfo {author} {\bibfnamefont {T.}~\bibnamefont {Best}}, \bibinfo {author}
  {\bibfnamefont {I.}~\bibnamefont {Bloch}}, \bibinfo {author} {\bibfnamefont
  {T.~A.}\ \bibnamefont {Costi}}, \bibinfo {author} {\bibfnamefont {R.~W.}\
  \bibnamefont {Helmes}}, \bibinfo {author} {\bibfnamefont {D.}~\bibnamefont
  {Rasch}},\ and\ \bibinfo {author} {\bibfnamefont {A.}~\bibnamefont {Rosch}},\
  }\bibfield  {title} {\bibinfo {title} {{Metallic and Insulating Phases of
  Repulsively Interacting Fermions in a 3D Optical Lattice}},\ }\href
  {https://doi.org/10.1126/science.1165449} {\bibfield  {journal} {\bibinfo
  {journal} {Science}\ }\textbf {\bibinfo {volume} {322}},\ \bibinfo {pages}
  {1520} (\bibinfo {year} {2008})}\BibitemShut {NoStop}%
\bibitem [{\citenamefont {J{\"{o}}rdens}\ \emph {et~al.}(2008)\citenamefont
  {J{\"{o}}rdens}, \citenamefont {Strohmaier}, \citenamefont {G{\"{u}}nter},
  \citenamefont {Moritz},\ and\ \citenamefont {Esslinger}}]{Jordens2008}%
  \BibitemOpen
  \bibfield  {author} {\bibinfo {author} {\bibfnamefont {R.}~\bibnamefont
  {J{\"{o}}rdens}}, \bibinfo {author} {\bibfnamefont {N.}~\bibnamefont
  {Strohmaier}}, \bibinfo {author} {\bibfnamefont {K.}~\bibnamefont
  {G{\"{u}}nter}}, \bibinfo {author} {\bibfnamefont {H.}~\bibnamefont
  {Moritz}},\ and\ \bibinfo {author} {\bibfnamefont {T.}~\bibnamefont
  {Esslinger}},\ }\bibfield  {title} {\bibinfo {title} {{A Mott insulator of
  fermionic atoms in an optical lattice}},\ }\href
  {https://doi.org/10.1038/nature07244} {\bibfield  {journal} {\bibinfo
  {journal} {Nature}\ }\textbf {\bibinfo {volume} {455}},\ \bibinfo {pages}
  {204} (\bibinfo {year} {2008})}\BibitemShut {NoStop}%
\bibitem [{\citenamefont {Gemelke}\ \emph {et~al.}(2009)\citenamefont
  {Gemelke}, \citenamefont {Zhang}, \citenamefont {Hung},\ and\ \citenamefont
  {Chin}}]{Gemelke2009}%
  \BibitemOpen
  \bibfield  {author} {\bibinfo {author} {\bibfnamefont {N.}~\bibnamefont
  {Gemelke}}, \bibinfo {author} {\bibfnamefont {X.}~\bibnamefont {Zhang}},
  \bibinfo {author} {\bibfnamefont {C.~L.}\ \bibnamefont {Hung}},\ and\
  \bibinfo {author} {\bibfnamefont {C.}~\bibnamefont {Chin}},\ }\bibfield
  {title} {\bibinfo {title} {{In situ observation of incompressible
  Mott-insulating domains in ultracold atomic gases}},\ }\href
  {https://doi.org/10.1038/nature08244} {\bibfield  {journal} {\bibinfo
  {journal} {Nature}\ }\textbf {\bibinfo {volume} {460}},\ \bibinfo {pages}
  {995} (\bibinfo {year} {2009})}\BibitemShut {NoStop}%
\bibitem [{\citenamefont {Greif}\ \emph {et~al.}(2016)\citenamefont {Greif},
  \citenamefont {Parsons}, \citenamefont {Mazurenko}, \citenamefont {Chiu},
  \citenamefont {Blatt}, \citenamefont {Huber}, \citenamefont {Ji},\ and\
  \citenamefont {Greiner}}]{Greif2016}%
  \BibitemOpen
  \bibfield  {author} {\bibinfo {author} {\bibfnamefont {D.}~\bibnamefont
  {Greif}}, \bibinfo {author} {\bibfnamefont {M.~F.}\ \bibnamefont {Parsons}},
  \bibinfo {author} {\bibfnamefont {A.}~\bibnamefont {Mazurenko}}, \bibinfo
  {author} {\bibfnamefont {C.~S.}\ \bibnamefont {Chiu}}, \bibinfo {author}
  {\bibfnamefont {S.}~\bibnamefont {Blatt}}, \bibinfo {author} {\bibfnamefont
  {F.}~\bibnamefont {Huber}}, \bibinfo {author} {\bibfnamefont
  {G.}~\bibnamefont {Ji}},\ and\ \bibinfo {author} {\bibfnamefont
  {M.}~\bibnamefont {Greiner}},\ }\bibfield  {title} {\bibinfo {title}
  {{Site-resolved imaging of a fermionic Mott insulator}},\ }\href
  {https://doi.org/10.1126/science.aad9041} {\bibfield  {journal} {\bibinfo
  {journal} {Science}\ }\textbf {\bibinfo {volume} {351}},\ \bibinfo {pages}
  {953} (\bibinfo {year} {2016})}\BibitemShut {NoStop}%
\bibitem [{\citenamefont {Bloch}\ \emph {et~al.}(2008)\citenamefont {Bloch},
  \citenamefont {Dalibard},\ and\ \citenamefont {Zwerger}}]{Bloch2008}%
  \BibitemOpen
  \bibfield  {author} {\bibinfo {author} {\bibfnamefont {I.}~\bibnamefont
  {Bloch}}, \bibinfo {author} {\bibfnamefont {J.}~\bibnamefont {Dalibard}},\
  and\ \bibinfo {author} {\bibfnamefont {W.}~\bibnamefont {Zwerger}},\
  }\bibfield  {title} {\bibinfo {title} {{Many-body physics with ultracold
  gases}},\ }\href {https://doi.org/10.1103/RevModPhys.80.885} {\bibfield
  {journal} {\bibinfo  {journal} {Reviews of Modern Physics}\ }\textbf
  {\bibinfo {volume} {80}},\ \bibinfo {pages} {885} (\bibinfo {year}
  {2008})}\BibitemShut {NoStop}%
\bibitem [{\citenamefont {Bloch}\ \emph {et~al.}(2012)\citenamefont {Bloch},
  \citenamefont {Dalibard},\ and\ \citenamefont
  {Nascimb{\`{e}}ne}}]{Bloch2012}%
  \BibitemOpen
  \bibfield  {author} {\bibinfo {author} {\bibfnamefont {I.}~\bibnamefont
  {Bloch}}, \bibinfo {author} {\bibfnamefont {J.}~\bibnamefont {Dalibard}},\
  and\ \bibinfo {author} {\bibfnamefont {S.}~\bibnamefont {Nascimb{\`{e}}ne}},\
  }\bibfield  {title} {\bibinfo {title} {{Quantum simulations with ultracold
  quantum gases}},\ }\href {https://doi.org/10.1038/nphys2259} {\bibfield
  {journal} {\bibinfo  {journal} {Nature Physics}\ }\textbf {\bibinfo {volume}
  {8}},\ \bibinfo {pages} {267} (\bibinfo {year} {2012})}\BibitemShut {NoStop}%
\bibitem [{\citenamefont {Gross}\ and\ \citenamefont
  {Bloch}(2017)}]{Gross2017}%
  \BibitemOpen
  \bibfield  {author} {\bibinfo {author} {\bibfnamefont {C.}~\bibnamefont
  {Gross}}\ and\ \bibinfo {author} {\bibfnamefont {I.}~\bibnamefont {Bloch}},\
  }\bibfield  {title} {\bibinfo {title} {{Quantum simulations with ultracold
  atoms in optical lattices}},\ }\href
  {https://doi.org/10.1126/science.aal3837} {\bibfield  {journal} {\bibinfo
  {journal} {Science}\ }\textbf {\bibinfo {volume} {357}},\ \bibinfo {pages}
  {995} (\bibinfo {year} {2017})}\BibitemShut {NoStop}%
\bibitem [{\citenamefont {Stellmer}\ \emph {et~al.}(2009)\citenamefont
  {Stellmer}, \citenamefont {Tey}, \citenamefont {Huang}, \citenamefont
  {Grimm},\ and\ \citenamefont {Schreck}}]{Stellmer2009}%
  \BibitemOpen
  \bibfield  {author} {\bibinfo {author} {\bibfnamefont {S.}~\bibnamefont
  {Stellmer}}, \bibinfo {author} {\bibfnamefont {M.~K.}\ \bibnamefont {Tey}},
  \bibinfo {author} {\bibfnamefont {B.}~\bibnamefont {Huang}}, \bibinfo
  {author} {\bibfnamefont {R.}~\bibnamefont {Grimm}},\ and\ \bibinfo {author}
  {\bibfnamefont {F.}~\bibnamefont {Schreck}},\ }\bibfield  {title} {\bibinfo
  {title} {{Bose-Einstein Condensation of Strontium}},\ }\href
  {https://doi.org/10.1103/PhysRevLett.103.200401} {\bibfield  {journal}
  {\bibinfo  {journal} {Physical Review Letters}\ }\textbf {\bibinfo {volume}
  {103}},\ \bibinfo {pages} {200401} (\bibinfo {year} {2009})}\BibitemShut
  {NoStop}%
\bibitem [{\citenamefont {Stellmer}\ \emph {et~al.}(2013)\citenamefont
  {Stellmer}, \citenamefont {Grimm},\ and\ \citenamefont
  {Schreck}}]{Stellmer2013}%
  \BibitemOpen
  \bibfield  {author} {\bibinfo {author} {\bibfnamefont {S.}~\bibnamefont
  {Stellmer}}, \bibinfo {author} {\bibfnamefont {R.}~\bibnamefont {Grimm}},\
  and\ \bibinfo {author} {\bibfnamefont {F.}~\bibnamefont {Schreck}},\
  }\bibfield  {title} {\bibinfo {title} {Production of quantum-degenerate
  strontium gases},\ }\href {https://doi.org/10.1103/PhysRevA.87.013611}
  {\bibfield  {journal} {\bibinfo  {journal} {Phys. Rev. A}\ }\textbf {\bibinfo
  {volume} {87}},\ \bibinfo {pages} {013611} (\bibinfo {year}
  {2013})}\BibitemShut {NoStop}%
\bibitem [{\citenamefont {Ferrari}\ \emph {et~al.}(2006)\citenamefont
  {Ferrari}, \citenamefont {Poli}, \citenamefont {Sorrentino},\ and\
  \citenamefont {Tino}}]{Ferrari2006}%
  \BibitemOpen
  \bibfield  {author} {\bibinfo {author} {\bibfnamefont {G.}~\bibnamefont
  {Ferrari}}, \bibinfo {author} {\bibfnamefont {N.}~\bibnamefont {Poli}},
  \bibinfo {author} {\bibfnamefont {F.}~\bibnamefont {Sorrentino}},\ and\
  \bibinfo {author} {\bibfnamefont {G.~M.}\ \bibnamefont {Tino}},\ }\bibfield
  {title} {\bibinfo {title} {{Long-Lived Bloch Oscillations with Bosonic Sr
  Atoms and Application to Gravity Measurement at the Micrometer Scale}},\
  }\href {https://doi.org/10.1103/PhysRevLett.97.060402} {\bibfield  {journal}
  {\bibinfo  {journal} {Physical Review Letters}\ }\textbf {\bibinfo {volume}
  {97}},\ \bibinfo {pages} {060402} (\bibinfo {year} {2006})}\BibitemShut
  {NoStop}%
\bibitem [{\citenamefont {Lubasch}\ \emph {et~al.}(2011)\citenamefont
  {Lubasch}, \citenamefont {Murg}, \citenamefont {Schneider}, \citenamefont
  {Cirac},\ and\ \citenamefont {Ba{\~{n}}uls}}]{Lubasch2011}%
  \BibitemOpen
  \bibfield  {author} {\bibinfo {author} {\bibfnamefont {M.}~\bibnamefont
  {Lubasch}}, \bibinfo {author} {\bibfnamefont {V.}~\bibnamefont {Murg}},
  \bibinfo {author} {\bibfnamefont {U.}~\bibnamefont {Schneider}}, \bibinfo
  {author} {\bibfnamefont {J.~I.}\ \bibnamefont {Cirac}},\ and\ \bibinfo
  {author} {\bibfnamefont {M.-C.}\ \bibnamefont {Ba{\~{n}}uls}},\ }\bibfield
  {title} {\bibinfo {title} {{Adiabatic Preparation of a Heisenberg
  Antiferromagnet Using an Optical Superlattice}},\ }\href
  {https://doi.org/10.1103/PhysRevLett.107.165301} {\bibfield  {journal}
  {\bibinfo  {journal} {Physical Review Letters}\ }\textbf {\bibinfo {volume}
  {107}},\ \bibinfo {pages} {165301} (\bibinfo {year} {2011})}\BibitemShut
  {NoStop}%
\bibitem [{Note3()}]{Note3}%
  \BibitemOpen
  \bibinfo {note} {The laser configuration we show in \protect \cref
  {figimplementationsetup} is not the only option for realizing a lattice which
  has nodes at all array positions and antinodes at the impurity positions.
  Such a lattice can also be generated by aligning lasers with frequencies
  $\omega >\pi /a$ with a suitable angle between the beams.}\BibitemShut
  {Stop}%
\bibitem [{\citenamefont {G{\"{u}}nter}\ \emph {et~al.}(2006)\citenamefont
  {G{\"{u}}nter}, \citenamefont {St{\"{o}}ferle}, \citenamefont {Moritz},
  \citenamefont {K{\"{o}}hl},\ and\ \citenamefont {Esslinger}}]{Gunter2006}%
  \BibitemOpen
  \bibfield  {author} {\bibinfo {author} {\bibfnamefont {K.}~\bibnamefont
  {G{\"{u}}nter}}, \bibinfo {author} {\bibfnamefont {T.}~\bibnamefont
  {St{\"{o}}ferle}}, \bibinfo {author} {\bibfnamefont {H.}~\bibnamefont
  {Moritz}}, \bibinfo {author} {\bibfnamefont {M.}~\bibnamefont {K{\"{o}}hl}},\
  and\ \bibinfo {author} {\bibfnamefont {T.}~\bibnamefont {Esslinger}},\
  }\bibfield  {title} {\bibinfo {title} {{Bose-Fermi Mixtures in a
  Three-Dimensional Optical Lattice}},\ }\href
  {https://doi.org/10.1103/PhysRevLett.96.180402} {\bibfield  {journal}
  {\bibinfo  {journal} {Physical Review Letters}\ }\textbf {\bibinfo {volume}
  {96}},\ \bibinfo {pages} {180402} (\bibinfo {year} {2006})}\BibitemShut
  {NoStop}%
\bibitem [{\citenamefont {Ospelkaus}\ \emph {et~al.}(2006)\citenamefont
  {Ospelkaus}, \citenamefont {Ospelkaus}, \citenamefont {Wille}, \citenamefont
  {Succo}, \citenamefont {Ernst}, \citenamefont {Sengstock},\ and\
  \citenamefont {Bongs}}]{Ospelkaus2006}%
  \BibitemOpen
  \bibfield  {author} {\bibinfo {author} {\bibfnamefont {S.}~\bibnamefont
  {Ospelkaus}}, \bibinfo {author} {\bibfnamefont {C.}~\bibnamefont
  {Ospelkaus}}, \bibinfo {author} {\bibfnamefont {O.}~\bibnamefont {Wille}},
  \bibinfo {author} {\bibfnamefont {M.}~\bibnamefont {Succo}}, \bibinfo
  {author} {\bibfnamefont {P.}~\bibnamefont {Ernst}}, \bibinfo {author}
  {\bibfnamefont {K.}~\bibnamefont {Sengstock}},\ and\ \bibinfo {author}
  {\bibfnamefont {K.}~\bibnamefont {Bongs}},\ }\bibfield  {title} {\bibinfo
  {title} {{Localization of Bosonic Atoms by Fermionic Impurities in a
  Three-Dimensional Optical Lattice}},\ }\href
  {https://doi.org/10.1103/PhysRevLett.96.180403} {\bibfield  {journal}
  {\bibinfo  {journal} {Physical Review Letters}\ }\textbf {\bibinfo {volume}
  {96}},\ \bibinfo {pages} {180403} (\bibinfo {year} {2006})}\BibitemShut
  {NoStop}%
\bibitem [{\citenamefont {Catani}\ \emph {et~al.}(2008)\citenamefont {Catani},
  \citenamefont {{De Sarlo}}, \citenamefont {Barontini}, \citenamefont
  {Minardi},\ and\ \citenamefont {Inguscio}}]{Catani2008}%
  \BibitemOpen
  \bibfield  {author} {\bibinfo {author} {\bibfnamefont {J.}~\bibnamefont
  {Catani}}, \bibinfo {author} {\bibfnamefont {L.}~\bibnamefont {{De Sarlo}}},
  \bibinfo {author} {\bibfnamefont {G.}~\bibnamefont {Barontini}}, \bibinfo
  {author} {\bibfnamefont {F.}~\bibnamefont {Minardi}},\ and\ \bibinfo {author}
  {\bibfnamefont {M.}~\bibnamefont {Inguscio}},\ }\bibfield  {title} {\bibinfo
  {title} {{Degenerate Bose-Bose mixture in a three-dimensional optical
  lattice}},\ }\href {https://doi.org/10.1103/PhysRevA.77.011603} {\bibfield
  {journal} {\bibinfo  {journal} {Physical Review A}\ }\textbf {\bibinfo
  {volume} {77}},\ \bibinfo {pages} {011603} (\bibinfo {year}
  {2008})}\BibitemShut {NoStop}%
\bibitem [{\citenamefont {Heinz}\ \emph {et~al.}(2020)\citenamefont {Heinz},
  \citenamefont {Park}, \citenamefont {{\v{S}}anti{\'{c}}}, \citenamefont
  {Trautmann}, \citenamefont {Porsev}, \citenamefont {Safronova}, \citenamefont
  {Bloch},\ and\ \citenamefont {Blatt}}]{Heinz2020a}%
  \BibitemOpen
  \bibfield  {author} {\bibinfo {author} {\bibfnamefont {A.}~\bibnamefont
  {Heinz}}, \bibinfo {author} {\bibfnamefont {A.~J.}\ \bibnamefont {Park}},
  \bibinfo {author} {\bibfnamefont {N.}~\bibnamefont {{\v{S}}anti{\'{c}}}},
  \bibinfo {author} {\bibfnamefont {J.}~\bibnamefont {Trautmann}}, \bibinfo
  {author} {\bibfnamefont {S.~G.}\ \bibnamefont {Porsev}}, \bibinfo {author}
  {\bibfnamefont {M.~S.}\ \bibnamefont {Safronova}}, \bibinfo {author}
  {\bibfnamefont {I.}~\bibnamefont {Bloch}},\ and\ \bibinfo {author}
  {\bibfnamefont {S.}~\bibnamefont {Blatt}},\ }\bibfield  {title} {\bibinfo
  {title} {{State-Dependent Optical Lattices for the Strontium Optical
  Qubit}},\ }\href {https://doi.org/10.1103/PhysRevLett.124.203201} {\bibfield
  {journal} {\bibinfo  {journal} {Physical Review Letters}\ }\textbf {\bibinfo
  {volume} {124}},\ \bibinfo {pages} {203201} (\bibinfo {year}
  {2020})}\BibitemShut {NoStop}%
\bibitem [{\citenamefont {Endres}\ \emph {et~al.}(2016)\citenamefont {Endres},
  \citenamefont {Bernien}, \citenamefont {Keesling}, \citenamefont {Levine},
  \citenamefont {Anschuetz}, \citenamefont {Krajenbrink}, \citenamefont
  {Senko}, \citenamefont {Vuletic}, \citenamefont {Greiner},\ and\
  \citenamefont {Lukin}}]{Endres2016}%
  \BibitemOpen
  \bibfield  {author} {\bibinfo {author} {\bibfnamefont {M.}~\bibnamefont
  {Endres}}, \bibinfo {author} {\bibfnamefont {H.}~\bibnamefont {Bernien}},
  \bibinfo {author} {\bibfnamefont {A.}~\bibnamefont {Keesling}}, \bibinfo
  {author} {\bibfnamefont {H.}~\bibnamefont {Levine}}, \bibinfo {author}
  {\bibfnamefont {E.~R.}\ \bibnamefont {Anschuetz}}, \bibinfo {author}
  {\bibfnamefont {A.}~\bibnamefont {Krajenbrink}}, \bibinfo {author}
  {\bibfnamefont {C.}~\bibnamefont {Senko}}, \bibinfo {author} {\bibfnamefont
  {V.}~\bibnamefont {Vuletic}}, \bibinfo {author} {\bibfnamefont
  {M.}~\bibnamefont {Greiner}},\ and\ \bibinfo {author} {\bibfnamefont {M.~D.}\
  \bibnamefont {Lukin}},\ }\bibfield  {title} {\bibinfo {title} {{Atom-by-atom
  assembly of defect-free one-dimensional cold atom arrays}},\ }\href
  {https://doi.org/10.1126/science.aah3752} {\bibfield  {journal} {\bibinfo
  {journal} {Science}\ }\textbf {\bibinfo {volume} {354}},\ \bibinfo {pages}
  {1024} (\bibinfo {year} {2016})}\BibitemShut {NoStop}%
\bibitem [{\citenamefont {Hu}\ \emph {et~al.}(2017)\citenamefont {Hu},
  \citenamefont {Urvoy}, \citenamefont {Vendeiro}, \citenamefont
  {Cr{\'{e}}pel}, \citenamefont {Chen},\ and\ \citenamefont
  {Vuleti{\'{c}}}}]{Hu2017}%
  \BibitemOpen
  \bibfield  {author} {\bibinfo {author} {\bibfnamefont {J.}~\bibnamefont
  {Hu}}, \bibinfo {author} {\bibfnamefont {A.}~\bibnamefont {Urvoy}}, \bibinfo
  {author} {\bibfnamefont {Z.}~\bibnamefont {Vendeiro}}, \bibinfo {author}
  {\bibfnamefont {V.}~\bibnamefont {Cr{\'{e}}pel}}, \bibinfo {author}
  {\bibfnamefont {W.}~\bibnamefont {Chen}},\ and\ \bibinfo {author}
  {\bibfnamefont {V.}~\bibnamefont {Vuleti{\'{c}}}},\ }\bibfield  {title}
  {\bibinfo {title} {{Creation of a Bose-condensed gas of 87 Rb by laser
  cooling}},\ }\href {https://doi.org/10.1126/science.aan5614} {\bibfield
  {journal} {\bibinfo  {journal} {Science}\ }\textbf {\bibinfo {volume}
  {358}},\ \bibinfo {pages} {1078} (\bibinfo {year} {2017})}\BibitemShut
  {NoStop}%
\bibitem [{\citenamefont {Urvoy}\ \emph {et~al.}(2019)\citenamefont {Urvoy},
  \citenamefont {Vendeiro}, \citenamefont {Ramette}, \citenamefont
  {Adiyatullin},\ and\ \citenamefont {Vuleti{\'{c}}}}]{Urvoy2019}%
  \BibitemOpen
  \bibfield  {author} {\bibinfo {author} {\bibfnamefont {A.}~\bibnamefont
  {Urvoy}}, \bibinfo {author} {\bibfnamefont {Z.}~\bibnamefont {Vendeiro}},
  \bibinfo {author} {\bibfnamefont {J.}~\bibnamefont {Ramette}}, \bibinfo
  {author} {\bibfnamefont {A.}~\bibnamefont {Adiyatullin}},\ and\ \bibinfo
  {author} {\bibfnamefont {V.}~\bibnamefont {Vuleti{\'{c}}}},\ }\bibfield
  {title} {\bibinfo {title} {{Direct Laser Cooling to Bose-Einstein
  Condensation in a Dipole Trap}},\ }\href
  {https://doi.org/10.1103/PhysRevLett.122.203202} {\bibfield  {journal}
  {\bibinfo  {journal} {Physical Review Letters}\ }\textbf {\bibinfo {volume}
  {122}},\ \bibinfo {pages} {203202} (\bibinfo {year} {2019})}\BibitemShut
  {NoStop}%
\bibitem [{\citenamefont {Porras}\ and\ \citenamefont
  {Cirac}(2008)}]{Porras2008}%
  \BibitemOpen
  \bibfield  {author} {\bibinfo {author} {\bibfnamefont {D.}~\bibnamefont
  {Porras}}\ and\ \bibinfo {author} {\bibfnamefont {J.~I.}\ \bibnamefont
  {Cirac}},\ }\bibfield  {title} {\bibinfo {title} {{Collective generation of
  quantum states of light by entangled atoms}},\ }\href
  {https://doi.org/10.1103/PhysRevA.78.053816} {\bibfield  {journal} {\bibinfo
  {journal} {Physical Review A}\ }\textbf {\bibinfo {volume} {78}},\ \bibinfo
  {pages} {053816} (\bibinfo {year} {2008})}\BibitemShut {NoStop}%
\bibitem [{\citenamefont {Fukuhara}\ \emph {et~al.}(2009)\citenamefont
  {Fukuhara}, \citenamefont {Sugawa}, \citenamefont {Sugimoto}, \citenamefont
  {Taie},\ and\ \citenamefont {Takahashi}}]{Fukuhara2009}%
  \BibitemOpen
  \bibfield  {author} {\bibinfo {author} {\bibfnamefont {T.}~\bibnamefont
  {Fukuhara}}, \bibinfo {author} {\bibfnamefont {S.}~\bibnamefont {Sugawa}},
  \bibinfo {author} {\bibfnamefont {M.}~\bibnamefont {Sugimoto}}, \bibinfo
  {author} {\bibfnamefont {S.}~\bibnamefont {Taie}},\ and\ \bibinfo {author}
  {\bibfnamefont {Y.}~\bibnamefont {Takahashi}},\ }\bibfield  {title} {\bibinfo
  {title} {Mott insulator of ultracold alkaline-earth-metal-like atoms},\
  }\href {https://doi.org/10.1103/PhysRevA.79.041604} {\bibfield  {journal}
  {\bibinfo  {journal} {Phys. Rev. A}\ }\textbf {\bibinfo {volume} {79}},\
  \bibinfo {pages} {041604} (\bibinfo {year} {2009})}\BibitemShut {NoStop}%
\bibitem [{\citenamefont {Gonz{\'{a}}lez-Tudela}\ and\ \citenamefont
  {Cirac}(2018)}]{Gonzalez-Tudela2018}%
  \BibitemOpen
  \bibfield  {author} {\bibinfo {author} {\bibfnamefont {A.}~\bibnamefont
  {Gonz{\'{a}}lez-Tudela}}\ and\ \bibinfo {author} {\bibfnamefont {J.~I.}\
  \bibnamefont {Cirac}},\ }\bibfield  {title} {\bibinfo {title} {{Non-Markovian
  Quantum Optics with Three-Dimensional State-Dependent Optical Lattices}},\
  }\href {https://doi.org/10.22331/q-2018-10-01-97} {\bibfield  {journal}
  {\bibinfo  {journal} {Quantum}\ }\textbf {\bibinfo {volume} {2}},\ \bibinfo
  {pages} {97} (\bibinfo {year} {2018})}\BibitemShut {NoStop}%
\bibitem [{\citenamefont {{Dalla Torre}}\ \emph {et~al.}(2006)\citenamefont
  {{Dalla Torre}}, \citenamefont {Berg},\ and\ \citenamefont
  {Altman}}]{DallaTorre2006}%
  \BibitemOpen
  \bibfield  {author} {\bibinfo {author} {\bibfnamefont {E.~G.}\ \bibnamefont
  {{Dalla Torre}}}, \bibinfo {author} {\bibfnamefont {E.}~\bibnamefont
  {Berg}},\ and\ \bibinfo {author} {\bibfnamefont {E.}~\bibnamefont {Altman}},\
  }\bibfield  {title} {\bibinfo {title} {{Hidden Order in 1D Bose
  Insulators}},\ }\href {https://doi.org/10.1103/PhysRevLett.97.260401}
  {\bibfield  {journal} {\bibinfo  {journal} {Physical Review Letters}\
  }\textbf {\bibinfo {volume} {97}},\ \bibinfo {pages} {260401} (\bibinfo
  {year} {2006})}\BibitemShut {NoStop}%
\bibitem [{\citenamefont {Yao}\ \emph {et~al.}(2012)\citenamefont {Yao},
  \citenamefont {Laumann}, \citenamefont {Gorshkov}, \citenamefont {Bennett},
  \citenamefont {Demler}, \citenamefont {Zoller},\ and\ \citenamefont
  {Lukin}}]{Yao2012}%
  \BibitemOpen
  \bibfield  {author} {\bibinfo {author} {\bibfnamefont {N.~Y.}\ \bibnamefont
  {Yao}}, \bibinfo {author} {\bibfnamefont {C.~R.}\ \bibnamefont {Laumann}},
  \bibinfo {author} {\bibfnamefont {A.~V.}\ \bibnamefont {Gorshkov}}, \bibinfo
  {author} {\bibfnamefont {S.~D.}\ \bibnamefont {Bennett}}, \bibinfo {author}
  {\bibfnamefont {E.}~\bibnamefont {Demler}}, \bibinfo {author} {\bibfnamefont
  {P.}~\bibnamefont {Zoller}},\ and\ \bibinfo {author} {\bibfnamefont {M.~D.}\
  \bibnamefont {Lukin}},\ }\bibfield  {title} {\bibinfo {title} {{Topological
  Flat Bands from Dipolar Spin Systems}},\ }\href
  {https://doi.org/10.1103/PhysRevLett.109.266804} {\bibfield  {journal}
  {\bibinfo  {journal} {Physical Review Letters}\ }\textbf {\bibinfo {volume}
  {109}},\ \bibinfo {pages} {266804} (\bibinfo {year} {2012})}\BibitemShut
  {NoStop}%
\bibitem [{\citenamefont {Yao}\ \emph {et~al.}(2013)\citenamefont {Yao},
  \citenamefont {Gorshkov}, \citenamefont {Laumann}, \citenamefont
  {L{\"{a}}uchli}, \citenamefont {Ye},\ and\ \citenamefont {Lukin}}]{Yao2013}%
  \BibitemOpen
  \bibfield  {author} {\bibinfo {author} {\bibfnamefont {N.~Y.}\ \bibnamefont
  {Yao}}, \bibinfo {author} {\bibfnamefont {A.~V.}\ \bibnamefont {Gorshkov}},
  \bibinfo {author} {\bibfnamefont {C.~R.}\ \bibnamefont {Laumann}}, \bibinfo
  {author} {\bibfnamefont {A.~M.}\ \bibnamefont {L{\"{a}}uchli}}, \bibinfo
  {author} {\bibfnamefont {J.}~\bibnamefont {Ye}},\ and\ \bibinfo {author}
  {\bibfnamefont {M.~D.}\ \bibnamefont {Lukin}},\ }\bibfield  {title} {\bibinfo
  {title} {{Realizing Fractional Chern Insulators in Dipolar Spin Systems}},\
  }\href {https://doi.org/10.1103/PhysRevLett.110.185302} {\bibfield  {journal}
  {\bibinfo  {journal} {Physical Review Letters}\ }\textbf {\bibinfo {volume}
  {110}},\ \bibinfo {pages} {185302} (\bibinfo {year} {2013})}\BibitemShut
  {NoStop}%
\bibitem [{\citenamefont {Manmana}\ \emph {et~al.}(2013)\citenamefont
  {Manmana}, \citenamefont {Stoudenmire}, \citenamefont {Hazzard},
  \citenamefont {Rey},\ and\ \citenamefont {Gorshkov}}]{Manmana2013}%
  \BibitemOpen
  \bibfield  {author} {\bibinfo {author} {\bibfnamefont {S.~R.}\ \bibnamefont
  {Manmana}}, \bibinfo {author} {\bibfnamefont {E.~M.}\ \bibnamefont
  {Stoudenmire}}, \bibinfo {author} {\bibfnamefont {K.~R.~A.}\ \bibnamefont
  {Hazzard}}, \bibinfo {author} {\bibfnamefont {A.~M.}\ \bibnamefont {Rey}},\
  and\ \bibinfo {author} {\bibfnamefont {A.~V.}\ \bibnamefont {Gorshkov}},\
  }\bibfield  {title} {\bibinfo {title} {{Topological phases in ultracold
  polar-molecule quantum magnets}},\ }\href
  {https://doi.org/10.1103/PhysRevB.87.081106} {\bibfield  {journal} {\bibinfo
  {journal} {Physical Review B}\ }\textbf {\bibinfo {volume} {87}},\ \bibinfo
  {pages} {081106} (\bibinfo {year} {2013})}\BibitemShut {NoStop}%
\bibitem [{\citenamefont {Gong}\ \emph {et~al.}(2016)\citenamefont {Gong},
  \citenamefont {Maghrebi}, \citenamefont {Hu}, \citenamefont {Foss-Feig},
  \citenamefont {Richerme}, \citenamefont {Monroe},\ and\ \citenamefont
  {Gorshkov}}]{Gong2016}%
  \BibitemOpen
  \bibfield  {author} {\bibinfo {author} {\bibfnamefont {Z.-X.}\ \bibnamefont
  {Gong}}, \bibinfo {author} {\bibfnamefont {M.~F.}\ \bibnamefont {Maghrebi}},
  \bibinfo {author} {\bibfnamefont {A.}~\bibnamefont {Hu}}, \bibinfo {author}
  {\bibfnamefont {M.}~\bibnamefont {Foss-Feig}}, \bibinfo {author}
  {\bibfnamefont {P.}~\bibnamefont {Richerme}}, \bibinfo {author}
  {\bibfnamefont {C.}~\bibnamefont {Monroe}},\ and\ \bibinfo {author}
  {\bibfnamefont {A.~V.}\ \bibnamefont {Gorshkov}},\ }\bibfield  {title}
  {\bibinfo {title} {{Kaleidoscope of quantum phases in a long-range
  interacting spin-1 chain}},\ }\href
  {https://doi.org/10.1103/PhysRevB.93.205115} {\bibfield  {journal} {\bibinfo
  {journal} {Physical Review B}\ }\textbf {\bibinfo {volume} {93}},\ \bibinfo
  {pages} {205115} (\bibinfo {year} {2016})}\BibitemShut {NoStop}%
\bibitem [{\citenamefont {Baier}\ \emph {et~al.}(2016)\citenamefont {Baier},
  \citenamefont {Mark}, \citenamefont {Petter}, \citenamefont {Aikawa},
  \citenamefont {Chomaz}, \citenamefont {Cai}, \citenamefont {Baranov},
  \citenamefont {Zoller},\ and\ \citenamefont {Ferlaino}}]{Baier2016}%
  \BibitemOpen
  \bibfield  {author} {\bibinfo {author} {\bibfnamefont {S.}~\bibnamefont
  {Baier}}, \bibinfo {author} {\bibfnamefont {M.~J.}\ \bibnamefont {Mark}},
  \bibinfo {author} {\bibfnamefont {D.}~\bibnamefont {Petter}}, \bibinfo
  {author} {\bibfnamefont {K.}~\bibnamefont {Aikawa}}, \bibinfo {author}
  {\bibfnamefont {L.}~\bibnamefont {Chomaz}}, \bibinfo {author} {\bibfnamefont
  {Z.}~\bibnamefont {Cai}}, \bibinfo {author} {\bibfnamefont {M.}~\bibnamefont
  {Baranov}}, \bibinfo {author} {\bibfnamefont {P.}~\bibnamefont {Zoller}},\
  and\ \bibinfo {author} {\bibfnamefont {F.}~\bibnamefont {Ferlaino}},\
  }\bibfield  {title} {\bibinfo {title} {{Extended Bose-Hubbard models with
  ultracold magnetic atoms}},\ }\href {https://doi.org/10.1126/science.aac9812}
  {\bibfield  {journal} {\bibinfo  {journal} {Science}\ }\textbf {\bibinfo
  {volume} {352}},\ \bibinfo {pages} {201} (\bibinfo {year}
  {2016})}\BibitemShut {NoStop}%
\bibitem [{\citenamefont {Landig}\ \emph {et~al.}(2016)\citenamefont {Landig},
  \citenamefont {Hruby}, \citenamefont {Dogra}, \citenamefont {Landini},
  \citenamefont {Mottl}, \citenamefont {Donner},\ and\ \citenamefont
  {Esslinger}}]{Landig2016}%
  \BibitemOpen
  \bibfield  {author} {\bibinfo {author} {\bibfnamefont {R.}~\bibnamefont
  {Landig}}, \bibinfo {author} {\bibfnamefont {L.}~\bibnamefont {Hruby}},
  \bibinfo {author} {\bibfnamefont {N.}~\bibnamefont {Dogra}}, \bibinfo
  {author} {\bibfnamefont {M.}~\bibnamefont {Landini}}, \bibinfo {author}
  {\bibfnamefont {R.}~\bibnamefont {Mottl}}, \bibinfo {author} {\bibfnamefont
  {T.}~\bibnamefont {Donner}},\ and\ \bibinfo {author} {\bibfnamefont
  {T.}~\bibnamefont {Esslinger}},\ }\bibfield  {title} {\bibinfo {title}
  {{Quantum phases from competing short- and long-range interactions in an
  optical lattice}},\ }\href {https://doi.org/10.1038/nature17409} {\bibfield
  {journal} {\bibinfo  {journal} {Nature}\ }\textbf {\bibinfo {volume} {532}},\
  \bibinfo {pages} {476} (\bibinfo {year} {2016})}\BibitemShut {NoStop}%
\bibitem [{\citenamefont {Sandvik}(2010)}]{Sandvik2010}%
  \BibitemOpen
  \bibfield  {author} {\bibinfo {author} {\bibfnamefont {A.~W.}\ \bibnamefont
  {Sandvik}},\ }\bibfield  {title} {\bibinfo {title} {{Ground States of a
  Frustrated Quantum Spin Chain with Long-Range Interactions}},\ }\href
  {https://doi.org/10.1103/PhysRevLett.104.137204} {\bibfield  {journal}
  {\bibinfo  {journal} {Physical Review Letters}\ }\textbf {\bibinfo {volume}
  {104}},\ \bibinfo {pages} {137204} (\bibinfo {year} {2010})}\BibitemShut
  {NoStop}%
\bibitem [{\citenamefont {Amir}\ \emph {et~al.}(2009)\citenamefont {Amir},
  \citenamefont {Oreg},\ and\ \citenamefont {Imry}}]{Amir2009}%
  \BibitemOpen
  \bibfield  {author} {\bibinfo {author} {\bibfnamefont {A.}~\bibnamefont
  {Amir}}, \bibinfo {author} {\bibfnamefont {Y.}~\bibnamefont {Oreg}},\ and\
  \bibinfo {author} {\bibfnamefont {Y.}~\bibnamefont {Imry}},\ }\bibfield
  {title} {\bibinfo {title} {Slow relaxations and aging in the electron
  glass},\ }\href {https://doi.org/10.1103/PhysRevLett.103.126403} {\bibfield
  {journal} {\bibinfo  {journal} {Phys. Rev. Lett.}\ }\textbf {\bibinfo
  {volume} {103}},\ \bibinfo {pages} {126403} (\bibinfo {year}
  {2009})}\BibitemShut {NoStop}%
\end{thebibliography}%

\end{document}